\documentclass[floatfix]{emulateapj}

\usepackage{amsmath}
\usepackage[colorlinks = true,linkcolor = blue, citecolor = blue]{hyperref}
\usepackage{natbib}
\usepackage{multirow}
\usepackage{placeins}

\shorttitle{How Dense is Your Gas?}

\begin{document}
\title{How Dense is Your Gas? \\ {\footnotesize On the Recoverability of LVG Model Parameters}}
\author{R. Tunnard and T.\,R. Greve}
\affil{Department of Physics and Astronomy, University College London, Gower Street, London WC1E 6BT, UK} \email{richard.tunnard.13@ucl.ac.uk}

\begin{abstract}
We explore the recoverability of gas physical conditions with the Large Velocity Gradient (LVG) model, using the public code RADEX and the molecules HCN and CO. Examining a wide parameter range with a series of models of increasing complexity we use both grid and Monte Carlo Markov Chain (MCMC) methods to recover the input conditions, and quantify the inherent and noise induced uncertainties in the model results. We find that even with the benefit of generous assumptions the LVG models struggle to recover any parameter better than to within half a dex, although we find no evidence of systemic offsets. Examining isotopologue lines we demonstrate that it is always preferable to model the isotopologue abundance ratio as a free parameter, due to large biases introduced in all other parameters when an incorrect ratio is assumed. Finally, we explore the effects of the background radiation temperature on CO and HCN line ratios, with an emphasis on the effect of the CMB at $z>4$, and show that while the effect on the line ratios is minor, the effect on the SLED peak is significant and that {the CO$(1-0)$ line luminosity to H$_2$ mass conversion factor} ($\alpha_{\rm CO}$) needs to be altered to account for the loss of contrast against the hotter CMB as redshift increases.
\end{abstract}


\section{Introduction}

One of the most successful and widely used tools for interpreting the physical conditions of gas traced by molecular rotational lines is the Large Velocity Gradient (LVG) model, where the radiative transfer and rotational level populations are decoupled through the introduction of a photon escape probability \citep{Sobolev1960,deJong1975,Goldreich1976,Weiss2005,radex2007}. This approach has two key strengths: {\bf 1)} it does not assume Local Thermodynamic Equilibrium (LTE), and {\bf 2)} it is fast, allowing for grids of models to be run. There are many implementations of the LVG model, although currently the most widely used in the millimetre and sub-millimetre regime is the publicly available LVG code, RADEX\footnote{\url{http://home.strw.leidenuniv.nl/~moldata/radex.html}} \citep{radex2007}. 

While far less {sophisticated} than Monte Carlo models such as RATRAN\footnote{\url{http://home.strw.leidenuniv.nl/~michiel/ratran/}} and LIME\footnote{\url{https://github.com/lime-rt/lime}} \citep{RATRAN, LIME}, which are better able to model radiative transfer in very optically thick environments as well as working in 3D, the much greater speed of LVG codes, and the fact that they do not require models (or assumptions) regarding the distribution of the gas, means that in many instances, especially extragalactic observations, LVG models are still the most practical non-LTE approach. RATRAN and LIME also include the ability to input non-uniform temperatures and densities; while this is in principle possible with the LVG approach there is, so far, no publicly available code providing these features.

One of the most widely used LVG codes is RADEX. When it was introduced, RADEX was recommended as a tool for interpreting molecular line ratios, primarily through diagnostic plots. However, since then, it and other LVG codes have become extensively used as general purpose tools for extracting gas conditions, especially in extragalactic targets e.g., \citet{Krips2008, Greve2009, Rangwala2011, Kamenetzky2014, Papadopoulos2014, Viti2014, Tunnard2015}, where the plethora of molecular lines now being observed by the state-of-the-art single dish telescopes and interferometer arrays are allowing for modelling of multiple line ratios, and therefore more complex physical models.

Despite the wide spread use of LVG models, there have been no published studies on the ability of LVG models to recover the true gas parameters. We present here such a study, using an unmodified version of RADEX to explore the precision and accuracy of LVG models as a means for recovering gas parameters.

\bigskip

\noindent As a topic of active research and great interest for the study of star formation over cosmic times the CO spectral line energy distributions (SLEDs) of star-forming galaxies have been extensively studied, including with sophisticated hydrodynamic modelling with radiative transfer post-processing \citep{Narayanan2009, Narayanan2010, Narayanan2014, Olsen2015}. However, while these papers present state of the art models of the galaxies and their molecular emission, the effect of the CMB temperature is given only a cursory discussion. In {cold} molecular environments {in} high redshift galaxies the effects of the CMB on the low$-J$ CO lines can become highly significant at $z>2$ where the CMB temperature begins to exceed the energy of the $J=1-0$ line (5.53\,K), similar to the suppression of molecular lines in local ultraluminous infrared galaxies (ULIRGs) \citep{Papadopoulos2010}, where the hot dust leads to a very significant reduction in the observed molecular flux. This in turn leads to apparently under-excited CO SLEDs unless corrected for the effects of dust. The effect of the CMB as a function of redshift has been modelled by \citet{daCunha2013} for single lines of sight for both dust and CO.

\bigskip

\noindent The outline of the paper is as follows. We describe the models, assumptions and techniques in Section \ref{sec:models}. We begin by demonstrating both the accuracy of grids and MCMCs with and without a 10\% observation uncertainty, and then show how $^{13}$C observations can improve the results. We then investigate whether it is better to assume a canonical molecular abundance or introduce the abundance as a free parameter, before repeating this for the [HCN]/[H$^{13}$CN] abundance ratio. We finally use two toy models to explore the potential effects of the CMB at high redshifts, including galaxy integrated effects. The results are presented in Section \ref{sec:results}, and discussed in Section \ref{sec:discussion}. We conclude in section \ref{sec:conclusions}.


\section{Models and Methods}\label{sec:models}

In order to test the recoverability of LVG parameters we adopt an idealised scenario in that we generate lines using RADEX, from which we then attempt to recover the input parameters. With modern millimetre and sub-millimetre line observations there is a ubiquitous absolute calibration uncertainty of $\sim10-20\%$ in both interferometric and single dish observations, and this uncertainty usually dominates the overall uncertainty in the line flux. We explore the recovery accuracy both with zero errors and where all lines are given an (optimistic) uncertainty of $10\%$.

{We note that in practice there are further significant potential sources of error, beyond the scope of this paper. In the case of single dish observations there are complications arising from an unknown source size and hence uncertain (and likely different for different molecular lines) beam coupling factors. While interferometers can alleviate these issues to some extent by resolving the source, they are in turn susceptible to spatial filtering, the resolving out of large scale emission, and to side lobe artefacts.}

\subsection{Generating Lines}

For all of our tests we generate artificial lines using RADEX and collision rates from the Leiden Lambda database \citep{radex2007,Dumouchel2010,Yang2010}. The specifics depend upon the investigation in question, but the features common to most of the models are:
\begin{itemize}
\item The physical free parameters are drawn from a uniform distribution in log space.
\item We restrict the thermal pressure, $P/k_{\rm B}$, of the models to be $<1\times10^8$\,K\,cm$^{-3}$. Higher thermal pressures are very unlikely/rare (even ULIRGs only exhibit thermal pressures $\sim10^7$\,K\,cm$^{-3}$: \citealp{Dopita2005}), but also RADEX tends to fail at high thermal pressures, biasing the results\footnote{The specific error seems to vary, but the result is that high thermal pressure inputs are recovered extremely poorly, with a $\sim4$\,dex underestimate of kinetic temperature and a $\sim2$ dex overestimate of density. There is also a region of high thermal pressure where all fluxes are zero.}.
\item The virial parameter, $K_{\rm vir}$\footnote{The virial parameter, $K_{\rm vir}$ is defined as \citep[e.g.,][]{Greve2009,Papadopoulos2012,Papadopoulos2014}:
\begin{equation}
K_{\rm vir} = \frac{{\rm d}v/{\rm d}r}{\left({\rm d}v/{\rm d}r\right)_{\rm vir}} =\frac{1.54}{\sqrt{\alpha}}\frac{{\rm d}v}{{\rm d}r}\left(\frac{n_{{\rm H}_2}}{1000{\rm \,cm}^{-3}}\right)^{-0.5}.
\end{equation}
The geometric coefficient $\alpha$ ranges from $1-2.5$ with an expectation value $\left<\sqrt{\alpha}\right>=1.312$. The velocity gradient, ${\rm d}v/{\rm d}r$, describes the change in line-of-sight velocity through the cloud, and is in units of km\,s$^{-1}$\,pc$^{-1}$.}, is set to unity (corresponding to virialised clouds), so that for a given $n_{{\rm H}_2}$ d$v/$d$r$ is determined, and not a free parameter. It is however kept as a free parameter when attempting to recover input parameters with the Monte Carlo Markov Chain (MCMC) models.
\end{itemize}

\subsection{Introducing Errors}
	
	We introduce errors on the lines produced by RADEX. The motivation is to emulate reality, where there is a true, unknown, value of the line brightness temperature, but the observed value may be larger or smaller, primarily due to calibration uncertainties. In these tests we know the true value, and generate the randomised values.
	
	For the tests including observation calibration uncertainty we randomise the flux of each line individually, such that the new, uncertain flux is drawn from a Normal distribution with mean equal to the artificial line brightness temperature and with a standard deviation equal to 0.1 times the artificial line brightness temperature. We calculate line ratios using randomised line brightnesses, propagating the uncertainties so that the uncertainty in the line ratios is $\sim14\%$. This is the uncertainty used when calculating $\chi^2$ in the grids and MCMC traces.
	
	In all but one of the cases we work entirely in line brightness temperature, the exception being when we plot SLED fluxes. This is a further idealisation, as it does not take into account the additional errors when observing galaxies due to variations in molecular excitation which can lead to source sizes varying as a function of line transition.

\subsection{Grids and MCMC}

We explore two methods for recovering parameters from line ratios using RADEX: grids and MCMCs. We first adopt the traditional grid based method before demonstrating the capabilities and limitations of an MCMC approach.

The grid method is, in principle, uniquely simple: a grid of input parameters is generated, and RADEX fluxes and line ratios generated for each point. Lines or line ratios are then fitted (minimum $\chi^2$) to the grid, and the best fit parameters thus recovered. Some authors extend this by generating a full likelihood grid \citep{Ward2003,Kamenetzky2011,Rangwala2011,Papadopoulos2014}. The grid must be trimmed of spurious points where either RADEX or the LVG framework breaks down (due to extremely high optical depths). This is accomplished with appropriate checks of the RADEX output when generating the grid.

\smallskip

\noindent We also use an MCMC model to recover the parameters, as in \citet{Tunnard2015b} and similar to \citet{Kamenetzky2014} who adopted a Nested Sampling approach for two phase modelling of CO emission. Instead of fitting to a discrete grid the MCMC explores the continuous parameter space, sampling the posterior probability distribution. This also allows for the modelling of much more complex parameter spaces than is easily achievable with the grid method. Furthermore, it has the advantage of naturally producing posterior probabilities and being trivial to marginalise. 

The results of the MCMC are a chain of parameter values (called a trace): for a converged MCMC using the Metropolis-Hastings sampler the results should oscillate about the most probable values (assuming a singly peaked posterior), so that the mean of each parameter over the trace corresponds to the most probably parameter \citep[e.g.,][]{Hanson2001}. We present both the means and the parameters corresponding to the minimum $\chi^2$ set of parameters.

The MCMC method is slower than grids for low dimensional parameter spaces, presenting a significant overhead when running for many realisations of the randomised line ratios. Its true power lies in the ability to simultaneously fit multiple species and gas phases, avoiding the bias inherent in fitting multiple gas phases sequentially or by eye \citep{Kamenetzky2014}. Here, we run all models for $5\times10^3$ steps with a burn in of $1\times10^3$ steps, based off preliminary tests on convergence rates. This is less than ideal, but necessary to be able to run a large number of trials in a reasonable time.

\bigskip

\noindent For the grid recovery models we generate $10^5$ trials, each with physical parameters drawn randomly from the parameter space and recorded, subject to the pressure upper limit. For the MCMC only 1000 trials are generated due to the much greater run time of the MCMC. Since we are not directly comparing the grid and MCMC methods this is not a significant concern.

\subsection{Model Descriptions}

We present four simple models for testing the precision and accuracy of the LVG approach: 
\begin{enumerate}
\item A general randomised model, which randomly selects parameters from uniform distributions in log space and uses a single molecular species.
\item As above, but using a molecular species and its $^{13}$C isotopologue lines. We explore how well we can recover the isotopologue abundance ratio, and the variation when using a single $^{13}$C line and when using four $^{13}$C lines. This model is only run with the MCMC.
\item A study of the effects of assuming an erroneous molecular abundance, relative to adopting the molecular abundance as a free parameter.
\item A study of the effects of assuming an erroneous isotopologue abundance ratio when including a single $^{13}$C isotopologue line with four main isotopologue lines, relative to adopting the isotopologue abundance ratio as a free parameter.
\end{enumerate}
and two investigations of the effects of background radiation:
\begin{enumerate}
\setcounter{enumi}{4}
\item An exploration of the effect of the background radiation blackbody temperature, $T_{\rm bg}$, on line emission of HCN and CO.
\item A toy model of the effects of varying CMB temperatures on galaxy integrated CO and HCN emission. 
\end{enumerate}
These models represent a wide range of conditions and provide essential estimates of the optimal accuracy of LVG models. All of these model are run with and without the $10\%$ randomisation errors to distinguish between inherent and random noise induced offsets and uncertainties.

\bigskip

\noindent For the LVG testing we fix $T_{\rm bg}=3$\,K. For all comparison grids and MCMCs we adopt the parameter ranges $3 \leq T_{\rm k} \leq 1000$ and $10^2 \leq n_{{\rm H}_2} \leq 10^8$. For the grids $K_{\rm vir}$ is set to unity, while for the MCMC runs we also adopt $0.1 \leq {\rm d}v/{\rm d}r \leq 1000$, $0.5 \leq K_{\rm vir} \leq 2.0$ and for the models including $^{13}$C isotopologues $10 \leq X_{^{12}{\rm C}}/ X_{^{13}{\rm C}} \leq 1000$. 

The MCMC models require initial positions to either be provided or randomly generated. In this work, we generate random initial conditions for each MCMC, with each parameter, $\theta_i$, drawn from from a uniform random distribution centred on the middle of the prior range and with a width 0.3 times the parameter prior range (in log space).

\subsection{Randomised, Single Species}

The general model explores randomised $T_{\rm k}$ and $n_{{\rm H}_2}$, with fixed molecular abundances\footnote{$X_{\rm HCN}=2\times10^{-8}$, $X_{\rm CO}=5\times10^{-5}$.}, $T_{\rm bg}=3$\,K and $K_{\rm vir}=1$. The model was run separately for HCN and CO, with and without the $10\%$ randomisation errors on the lines fluxes, and the input parameters were recovered using both grid and MCMC methods.

This model serves to explore the ability of LVG models to recover the true physical parameters under the simplest and most ideal conditions.

For HCN we use only the $J=1-0$ to $J=4-3$ transitions, as these are the most commonly available lines (although all four lines are  available in the literature for only a few sources). For CO we use the full SLED up to the $J=13-12$ transition, as is available for many galaxies but in particular those in the Herschel Comprehensive ULIRG Emission Survey (HerCULES, PI: van der Werf, \citealp{vanderWerf2010, Rosenberg2015, Kamenetzky2015}). However, in reality this full SLED covers a wide range of physical conditions: here they all sample a single parameter set. Therefore, the derived uncertainties are a lower limit on the true uncertainties when interpreting real observations. Indeed, fitting a single gas phase to the full CO SLED of a galaxy is almost never possible, and even less frequently reliable. We leave an investigation into the effects of reducing a continuum of gas phases across a galaxy to a two or three phase model to a future paper.

\subsection{Randomised, with Isotopologues}

This model is identical to the Randomised, Single Species case, except that we now only use HCN and include H$^{13}$CN lines. We only run these models with the MCMC as the grid becomes a cube and starts to take prohibitively long to generate and compare with the synthetic line ratios. The [HCN]/[H$^{13}$CN] abundance ratio is entered into the model as a free parameter, while we keep $X_{\rm HCN}=2\times10^{-8}$ for both line generation and parameter recovery.

We run these tests using both the complete H$^{13}$CN SLED from $J=1-0$ up to $J=4-3$ as well as just the $J=1-0$ line. It is rare enough to have a single H$^{13}$CN observation in a galaxy: to have four lines is unheard of. Nevertheless we explore the four line case to assess the potential benefit of obtaining additional lines.

\subsection{Assumed $X_{\rm HCN}$}

It is standard practice with LVG modelling to assume a canonical molecular abundance. As well as simply being frequently necessitated by a paucity of observed lines this has a physical justification in that LVG models are only sensitive to the combination $X_{\rm mol}\left({\rm d}v/{\rm d}r\right)^{-1}$, not to either individually. This is sometimes then used in the RADEX native form $N_{\rm mol}/\Delta v$. However, since 

\begin{equation}
N_{\rm mol} = X_{\rm mol}N_{{\rm H}_2} = \frac{3.09\times10^{18}\,X_{\rm mol}\,n_{{\rm H}_2}\,\Delta v}{{\rm d}v/{\rm d}r},
\end{equation}
fixing $N_{\rm mol}/\Delta v$ actually forces certain, tacit, assumptions about the virial state of the gas\footnote{The numerical prefactor is just the parsec to cm conversion.}. Due to these complications we present a simple exploration of whether model parameters are better recovered assuming a canonical, but potentially erroneous, $X_{\rm mol}$, or whether better results are obtained by leaving $X_{\rm mol}$ as a free parameter.

Note that we do not claim that the molecular abundance can be recovered accurately when only a single species is observed. What we are investigating is whether the recovery of $T_{\rm k}$ and $n_{{\rm H}_2}$ is seriously affected by assuming a canonical $X_{\rm mol}$.

To this end we generate HCN lines with physical parameters drawn from the aforementioned parameter ranges, but we now also draw $X_{\rm HCN}$ from the range $10^{-12} - 10^{-4}$. However, when recovering the parameters with the MCMC we run two cases: the first where $X_{\rm HCN}$ is introduced as a free parameter, and the second where we fix $X_{\rm HCN}$ to the canonical value of $2\times10^{-8}$.

\subsection{Assumed [HCN]/[H$^{13}$CN]}

When only a single $^{13}$C isotopologue line is available it is common to assume a canonical molecular abundance and/or isotopologue abundance ratio. However, these values are poorly constrained for CO, let alone less abundant molecular species such as HCN. Furthermore, isotopologue selective chemical effects such as selective photodissociation and isotope charge exchange reactions can lead to isotope fractionation. When coupled with real variations in the elemental isotope abundances, such as in accreted primordial gas or cosmic ray dominated star formation paradigms, can lead to large variations in the isotopologue abundance ratios from species to species and galaxy to galaxy \citep{Henkel2010, Henkel2014,Papadopoulos2010b,Papadopoulos2011,Ritchey2011,Roueff2015}. We therefore explore the effects of assuming an incorrect abundance ratio on the recovered $T_{\rm k}$ and $n_{{\rm H}_2}$, using HCN. 

The [HCN]/[H$^{13}$CN] ratio is drawn from the range 10 to 1000 (as a uniform random variable in log-space), and we recover the parameters assuming [HCN]/[H$^{13}$CN]$ = 60$ and adopting a fixed $X_{\rm HCN}=2\times10^{-8}$\footnote{A case with both $X_{\rm HCN}$ and [HCN]/[H$^{13}$CN] as free parameters can be found in the Appendix.}. This allows us to test whether assuming a canonical value is reasonable, or whether it is in fact better to introduce [HCN]/[H$^{13}$CN] as a free parameter (as in the Randomised, with Isotopologues model), even when there is only a single H$^{13}$CN line available.

\subsection{$T_{\rm bg}$}

The exploration of the effects of $T_{\rm bg}$ on the lines is not a test of the abilities of LVG models to recover input parameters. Rather, it is a demonstration of the effects of background radiation on the molecular lines. We generate line fluxes for $T_{\rm bg} = 3$, 5, 10, 15, 20, 25 and 30\,K. For each $T_{\rm bg}$ we generate SLEDs from $J=1-0$ to $J=4-3$ for HCN and from $J=1-0$ to $J=13-12$ for CO, with constant density $n_{{\rm H}_2} = 1\times10^4$\,cm$^{-3}$ {(HCN)} and $1\times10^3$\,cm$^{-3}$ {(CO)}, and $X_{\rm HCN}=2\times10^{-8}$ {and} $X_{\rm CO}=5\times10^{-5}$, for $T_{\rm k} = 3$, 5, 10, 50, 100, 300, 500 and 1000\,K. For the CO lines we reduce the density to $1\times10^3$\,cm$^{-3}$ since at $1\times10^4$\,cm$^{-3}$ many lines, but in particular the $J=5-4$ line, present pathological fluxes (e.g., for $T_{\rm bg}=3$\,K and $T_{\rm k}\geq500$\,K RADEX fails to converge, ultimately because the column density and associated optical depth becomes too large). This is also more representative of the likely conditions of the CO on galactic scales. We also generate grids to explore the effect of $T_{\rm bg}$ on the line brightness ratios, with high resolution in $T_{\rm bg}$ and $T_{\rm k}$.

\bigskip

\noindent We are also interested in how the CO and HCN line peak brightness temperatures, $T_{\rm B, peak}$, are affected by the CMB, as a function of the kinetic temperature of the emitting gas. This allows us to investigate the detectability of cold gas reservoirs at high $z$. For five redshifts (0, 2, 4, 6 and 8) we run RADEX for the physical conditions above for 1000 values of $T_{\rm k}$ distributed evenly in log-space from 1000\,K to 3\,K. Other physical parameters are kept as for the models above.

\subsection{Toy Galaxy Models}\label{subsec:toys}

We extend the functional models above to a very simple 1D toy galaxy model in order to explore the effect of CMB temperature on potential galaxy integrated SLEDs.

The model uses a $1/r$ temperature profile falling from 100\,K at $r=0$ down to 10\,K at the galaxy edge, and we sample the profile with $10^4$ molecular clouds following an exponential distribution, and then integrate the total emission from all of the clouds in the galaxy. We do not explicitly include dust, but we do set $T_{\rm CMB}$ as a lower limit on $T_{\rm k}$. It is however still dust free, as RADEX does not include dust \emph{within} the gas: dust can only be added as a source of background radiation incident on the cloud. This toy model allows us to demonstrate the significant effects of the CMB on CO and HCN emission and on their line ratios at high $z$.

These toy models are much less complex than the work of \citet{daCunha2013} and \citet{Narayanan2014}, with an emphasis lying somewhere between the two. While \citet{daCunha2013} focussed heavily on the theory of line excitation as a function of redshift, they did not investigate galaxy wide trends. On the other hand, \citet{Narayanan2014} focussed almost entirely on detailed modelling of galaxy emission, with little investigation of the effect of the CMB at high redshift (although see their Figure 12). Our approach here is to simply demonstrate the possibility of unifying the two approaches.


\section{Results}\label{sec:results}

We present the results of the models in the following sections. For the sake of readability, for the recoverability tests we present the results in tables in the main text and include the figures in the Appendix. 

\subsection{Randomised, Single Species}

The results of the Randomised, Single Species models are presented in Figure \ref{fig:hcnco_grids} in the text and Figures \ref{fig:MCMChcn} and \ref{fig:MCMCco} in the Appendix, where we plot the histograms of the logarithm of the ratio of the recovered over true parameters (hereafter the ``deviation''). Numerical results are included in Table \ref{tab:1}. For the grid models we also plot the deviation of the product $T_{\rm k}\times n_{{\rm H}_2}$, which is more tightly constrained than $T_{\rm k}$ or $n_{{\rm H}_2}$ alone. We record the means and standard deviations of the deviations, {as well as the bootstrap estimated mean and standard deviation of the median of the deviations}.

\begin{table*}[h]
\centering
\caption{Deviations of Recovered Parameters from Grid Models}\label{tab:1}
\begin{tabular}{l c c c c c c}\hline
& \multicolumn{3}{c}{Statistical} & \multicolumn{3}{c}{Bootstrap Median}\\
& $T_{\rm k}$ & $n_{{\rm H}_2}$ & $T_{\rm k} \times n_{{\rm H}_2}$ & $T_{\rm k}$ & $n_{{\rm H}_2}$ & $T_{\rm k} \times n_{{\rm H}_2}$ \\[0.1cm] \hline\hline
HCN & $-0.2\pm0.5$ & $0.3\pm0.8$ & $0.1\pm0.5$ & $-0.085\pm0.001$ & $0.194\pm0.002$ & $0.058\pm0.001$ \\
HCN-noE & $-0.02\pm0.20$ & $0.05\pm0.37$ & $0.03\pm 0.29$ & $0.0001\pm0.0003$ & $0.0007\pm0.0003$ & $0.0038\pm0.0003$\\
CO & $-0.2\pm0.5$ & $0.2\pm0.9$ & $0.0\pm0.9$ & $-0.058\pm0.001$ & $0.071\pm0.001$ & $0.024\pm0.001$\\
CO-noE & $-0.1\pm0.4$ & $0.0\pm0.5$ & $-0.1\pm0.7$ & $-0.0120\pm0.0003$ & $0.0081\pm0.0004$ & $0.0013\pm0.0003$\\ \hline
\end{tabular}
\end{table*}

\begin{figure*}[h]
\centering
\includegraphics[width=\textwidth]{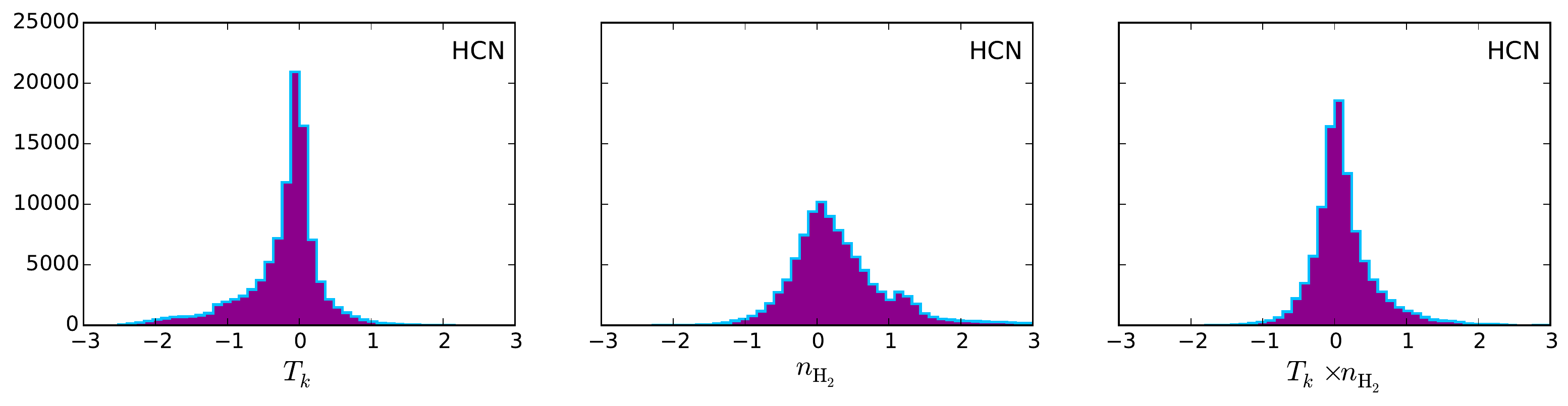}\\
\includegraphics[width=\textwidth]{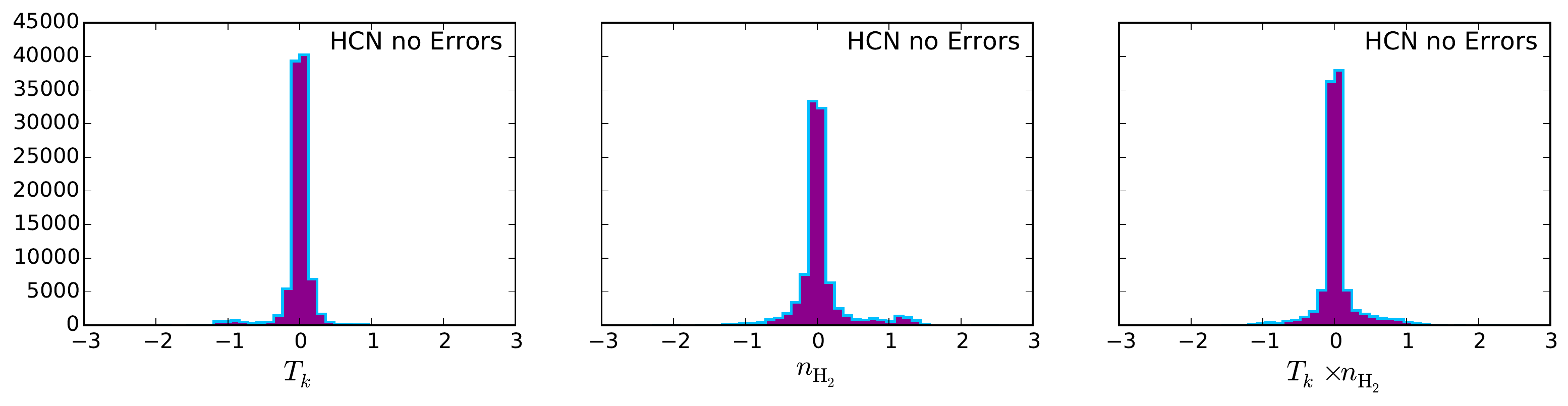}\\
\includegraphics[width=\textwidth]{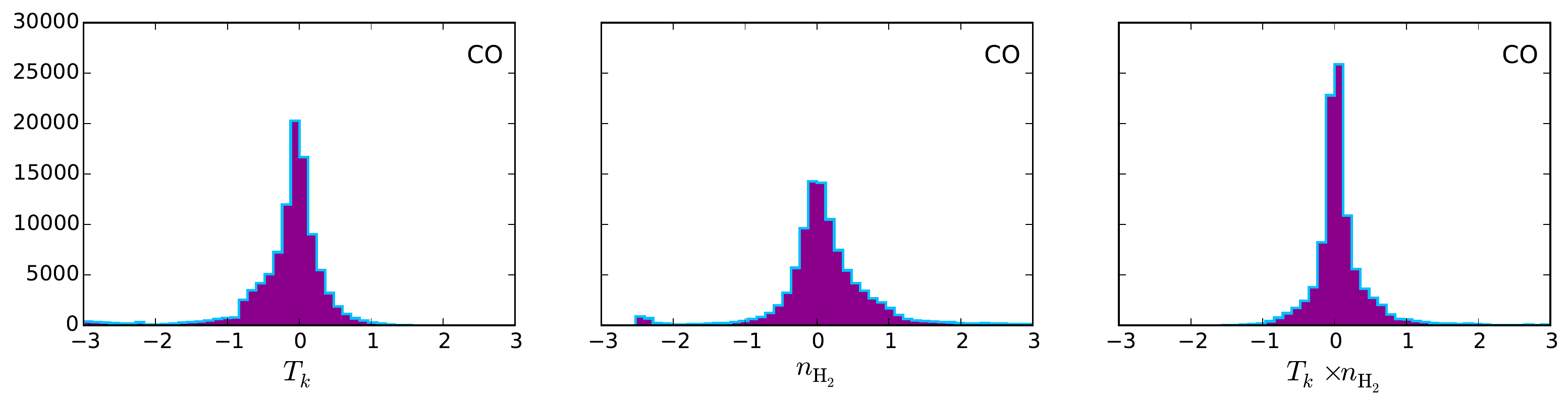}\\
\includegraphics[width=\textwidth]{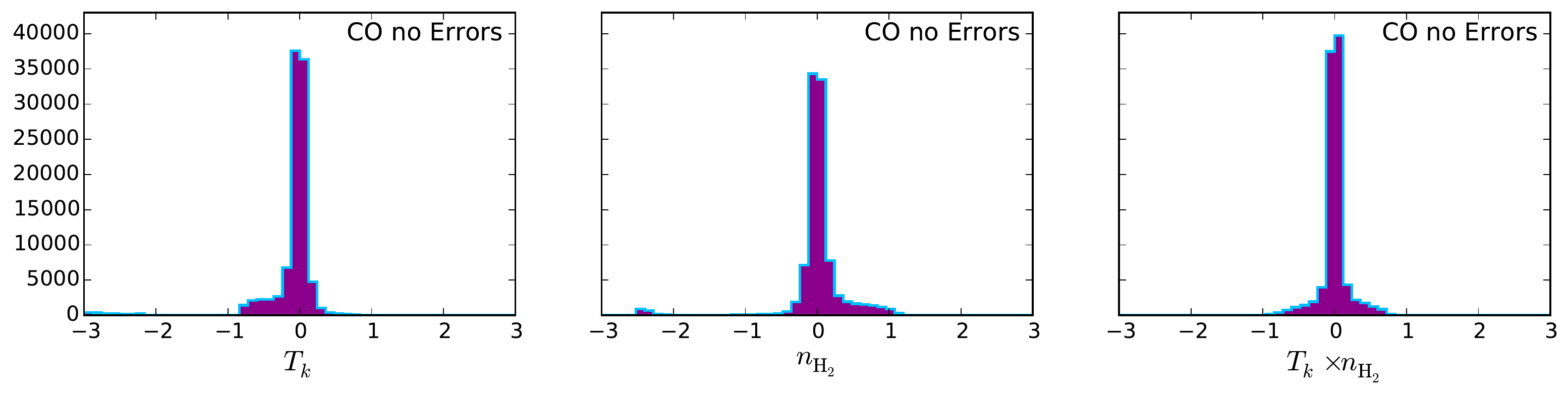}
\caption{The logarithm of the ratio of the recovered to true parameters for HCN and with and without $10\%$ errors as indicated in the upper right corner of each plot, drawn from a pressure restricted parameter range and recovered with a grid with a log step size of 0.1.}\label{fig:hcnco_grids}
\end{figure*}

\begin{table*}
\centering
\caption{Deviations of Recovered Parameters from MCMC Models.}\label{tab:2}
\begin{tabular}{l c c c c c c c}\hline
& \multicolumn{3}{c}{Statistical} & \multicolumn{3}{c}{Bootstrap Median}\\
& $T_{\rm k}$ & $n_{{\rm H}_2}$ & d$v/$d$r$ & $T_{\rm k}$ & $n_{{\rm H}_2}$ & d$v/$d$r$ &  \\[0.1cm] \hline\hline
HCN Best fit & $-0.2\pm0.6$ & $0.2\pm0.9$ & $0.1\pm0.6$ & $-0.013\pm0.003$ & $0.07\pm0.02$ & $0.07\pm0.02$ \\
HCN Mean & $-0.2\pm0.6$ & $0.2\pm1.0$ & $0.1\pm0.5$ & $-0.015\pm0.003$ & $0.04\pm0.01$ & $0.03\pm0.01$\\
HCN-noE Best fit & $-0.3\pm0.7$ & $0.2\pm1.0$ & $0.1\pm0.6$ & $-0.0010\pm0.0003$ & $0.004\pm0.002$ & $0.009\pm0.003$\\
HCN-noE Mean & $-0.3\pm0.7$ & $0.2\pm1.1$ & $0.1\pm0.6$ & $-0.004\pm0.001$ & $0.02\pm0.01$ & $0.031\pm0.009$\\ 
CO Best fit & $-0.5\pm0.9$ & $0.5\pm1.8$ & $0.3\pm0.9$ & $-0.010\pm0.003$ & $0.05\pm0.02$ & $0.11\pm0.03$ \\
CO Mean & $-0.5\pm0.9$ & $0.5\pm1.8$ & $0.3\pm0.9$ & $-0.011\pm0.003$ & $0.08\pm0.01$ & $0.11\pm0.02$\\ 
CO-noE Best fit & $-0.6\pm1.0$ & $1.0\pm2.0$ & $0.6\pm1.0$ &  $-0.005\pm0.001$ & $0.06\pm0.02$ & $0.09\pm0.03$\\
CO-noE Mean & $-0.6\pm1.0$ & $1.1\pm1.9$ & $0.6\pm1.0$ & $-0.018\pm0.005$ & $0.34\pm0.07$ & $0.29\pm0.03$ \\ \hline
\end{tabular}
\end{table*}

The numerical results for the grid models are shown in Table \ref{tab:1}. We include the statistical mean and standard deviations of the deviations, as well as the {bootstrap estimated} means and standard deviations of {the medians of the }deviations. Reassuringly, all parameters are recovered without bias for the no-error cases. There is a trend for both HCN and CO, when line uncertainties are added, to underestimate $T_{\rm k}$ and overestimate $n_{{\rm H}_2}$. This trend is {far weaker in the medians, suggesting that} it is due to the rare but extreme outliers. There is no sign that the CO model, with 9 more lines than the HCN model, obtained improved recovery rates above those of HCN.

The thermal pressure, $T_{\rm k}\times n_{{\rm H}_2}$, is better determined than either of $T_{\rm k}$ or $n_{{\rm H}_2}$ alone. The effects of increasing $T_{\rm k}$ and increasing $n_{{\rm H}_2}$ are not identical, but they are partially degenerate, leading to the improved recovery of thermal pressure.

The results show that the recovery rates are worryingly low. Without any errors there is a spread of $0.05-0.10$\,dex in all of the recovered parameters. I.e., even if we knew the line ratios perfectly, in a spatial resolved, virialised gas cloud of uniform density, temperature and velocity gradient, with a uniform and known molecular abundance, we should not expect to recover parameters with a precision better than $\sim10\%$. Keeping all other idealisations and introducing $10\%$ observational uncertainties this falls to between $50\%$ and $300\%$. These are lower limits on the true uncertainties. The accuracies are very good in the no error cases, but for the randomised cases the systematic offsets can be as large as a factor of 2.

\bigskip

\noindent The MCMC results are recorded in Table \ref{tab:2}. For the MCMC we have results for both the trace mean parameters and the minimum $\chi^2$ parameters. We also record the median of the standard deviations of the traces for each parameter, as these are one measure of the uncertainty in the MCMC mean results. The systematics seen in the grids are also present here, and once again it is readily apparent that they are due to outliers.

The MCMC results are particularly interesting. The MCMC appears to have much less Gaussian deviation distributions, being very strongly peaked about the true parameters, but with low levels of extreme outliers. For example, in the no error case in Figure \ref{fig:MCMCco} between $55\%$ and $68\%$ of the $T_{\rm k}$ values were recovered very nearly perfectly, but there is a low level tail of underestimates extending out to a factor of 300.

The mean parameters of the MCMC traces should be more meaningful than the best fit parameters. This is consistent with the d$v/$d$r$ deviations for the randomised MCMC model, where the best fit values almost form a top-hat distribution, while the mean values are very nearly Normally distributed.

The CO results are consistently worse than the HCN results, with a far greater tendency towards extremely low $T_{\rm k}$ and high $n_{{\rm H}_2}$. It is not clear why this is the case, although further examination revealed that the outliers all had very poor MCMC traces due to non-physical synthetic fluxes. These arose even when RADEX converged and optical depths were reasonable: there were specific parameter combinations which appear to produce pathological fluxes (as was found for some combinations in the $T_{\rm bg}$ exploration). Small variations away from the specific input parameters leads to the recovery of realistic SLEDs. While no method was found to reliably exclude these values from the synthetic fluxes we do not consider this of great concern: when fitting to real data the MCMC will find very poor fits for these values and so exclude them, jumping to a nearby point. Since the surrounding phase space will produce good fits this excludes only an N-dimensional delta-function from the trace, and the results are still reliable. This is supported by the \emph{inability} of the MCMC to fit these pathological synthetic fluxes: if the responsible phase-volume was non-zero the MCMC would still be able to ``home in'' on it. Therefore, this is only a problem for the synthetic fluxes and not for real observations. For the synthetic fluxes it produces the extreme outliers, which are clearly distinguishable post-hoc. The pseudo-Gaussian spread in the deviations is unrelated to this issue, and is due to many-to-one relationship between physical conditions and line ratios which is the focus of this work.

\bigskip

\noindent These grid and MCMC results demonstrate that even under ideal conditions a 10\% calibration uncertainty corresponds to uncertainties in LVG derived parameters of approximately 0.5\,dex / a factor of 3. This is believed to be due to both the slow variation of the line ratios across the parameter space and the many-to-one relationship between physical conditions and line ratios.

\subsection{Randomised, with Isotopologues}

The results of the Randomised, with Isotopologues trials with four H$^{13}$CN lines and one H$^{13}$CN line are shown in Figures \ref{fig:c12c13_4lines} and \ref{fig:c12c13_1line} respectively, and the numerical results are collated in Table \ref{tab:13C}.

The figures demonstrate the (well known) power of adding isotopologue lines. Even with the additional free parameter of [HCN]/[H$^{13}$CN] the model places much tighter and more accurate constraints on $T_{\rm k}$ and $n_{{\rm H}_2}$, as well as placing very tight constraints on [HCN]/[H$^{13}$CN]. This is not simply an effect of having more lines to work with since the fits are better than those obtained using the 13 CO lines. The good $T_{\rm k}$ results are due to the strong degeneracy between $T_{\rm k}$ and [HCN]/[H$^{13}$CN]. This can be roughly understood as follows: $T_{\rm k}$ and $n_{{\rm H}_2}$ set the primary shape of the SLEDs of HCN and H$^{13}$CN - with this determined the model can then find [HCN]/[H$^{13}$CN] since, to first order, this corresponds to a scaling of the SLED but not a change in shape\footnote{Although second order effects come into play due to the differing optical depths and line trapping.}. Put another way: iso-ratio contours for HCN and H$^{13}$CN intersect roughly perpendicularly in $T_{\rm k} - n_{{\rm H}_2}$ plots. None of this needs to be entered into the model: it explores the posterior probability space, without needing to understand the underlying relationships. With a single H$^{13}$CN line the ability to uniquely identify a $T_{\rm k}$ is lost, due again to the degeneracy between $T_{\rm k}$ and [HCN]/[H$^{13}$CN].

The four H$^{13}$CN lines, no error results are exceptionally good, with the models recovering the correct $T_{\rm k}$ and [HCN]/[H$^{13}$CN] in almost 100\% of the cases, and $n_{{\rm H}_2}$ and d$v/$d$r$ being exceptionally well constrained. There are however a few extreme outliers, which are responsible for the statistical deviations in Table \ref{tab:13C} being so much greater than {those of the medians}. The single H$^{13}$CN line fits are still very good; although worse than the 4 H$^{13}$CN line fits they are still better than the HCN only models, despite the additional free parameter.

\bigskip

\begin{table*}
\centering
\caption{Deviations of Recovered Parameters from MCMC $^{13}$C Isotopologue Models.\\Values in parenthesis are the MCMC standard deviations.}\label{tab:13C}
\begin{tabular}{c c c c c c}\hline
& & \multicolumn{2}{c}{Four H$^{13}$CN lines} & \multicolumn{2}{c}{One H$^{13}$CN line} \\ 
& & Best & Mean & Best & Mean \\ \hline\hline
\multirow{4}{*}{\rotatebox[origin=c]{90}{Statistical}} & $T_{\rm k}$ & $-0.1\pm0.4(0.04)$ & $-0.1\pm0.4(0.04)$ & $-0.1\pm0.5(0.1)$ & $-0.1\pm0.5(0.1)$\\
& $n_{{\rm H}_2}$ & $-0.0\pm0.7(0.14)$ & $-0.1\pm0.7(0.14)$ & $0.0\pm0.6(0.2)$ & $0.0\pm0.6(0.2)$\\
& ${\rm d}v/{\rm d}r$ & $0.0\pm0.4(0.18)$ & $0.0\pm0.4(0.18)$ & $0.0\pm0.4(0.2)$ & $0.0\pm0.3(0.2)$ \\
& $\frac{[{\rm HCN}]}{[{\rm H}^{13}{\rm CN}]}$ & $0.01\pm0.21(0.03)$ & $0.01\pm0.22(0.03)$ & $0.05\pm0.28(0.05)$ & $0.05\pm0.29(0.05)$ \\ \hline
\multirow{4}{*}{\rotatebox[origin=c]{90}{Median}} & $T_{\rm k}$ & $-0.003\pm0.002$ & $-0.005\pm0.002$ & $-0.006\pm0.003$ & $-0.010\pm0.003$\\
& $n_{{\rm H}_2}$ & $-0.019\pm0.015$ & $-0.025\pm0.010$ & $0.015\pm0.016$ & $-0.007\pm0.012$\\
& ${\rm d}v/{\rm d}r$ & $-0.063\pm0.016$ & $-0.034\pm0.009$ & $0.029\pm0.016$ & $-0.006\pm0.010$\\
& $\frac{[{\rm HCN}]}{[{\rm H}^{13}{\rm CN}]}$ & $0.002\pm0.002$ & $-0.001\pm0.003$ & $0.005\pm0.003$ & $0.009\pm0.003$\\ \hline 

\end{tabular}
\end{table*}

\noindent The value of multiple $^{13}$C isotopologue lines is clear. However, despite only being $10-15\times$ fainter than the $^{12}$C lines\footnote{Even when $100\times$ less abundant, due to optical depth effects.} the acquisition of even one H$^{13}$CN line in a local galaxy $(z\leq0.1)$ represents a significant time investment, even with ALMA. The small size of these sources, coupled with the inherent faintness of the $^{13}$C lines, leads to them being washed out by beam dilution, requiring even longer to observe with single dish instruments. The one exception to this is $^{13}$CO, which is sufficiently abundant to emit relatively brightly and be quite easily detectable. In cases where accurate and precise non-LTE gas parameters are desired from molecular observations it is imperative that $^{13}$C isotopologue lines be observed.

\subsection{Assumed $X_{\rm HCN}$}

The results of the third model, with free and fixed $X_{\rm HCN}$, are shown in Figure \ref{fig:fixXhcn} and Table \ref{tab:FreeHCN}. Note in particular the much wider ranges of the x-axes than in the previous plots. 

There is no clear sign of any difference in the recoverability of $T_{\rm k}$ or $n_{{\rm H}_2}$ in either model. Furthermore, the molecular abundance in the free parameter case is essentially unconstrained and although the model is slightly able to recover the true value there is a huge spread in the results as well as a systematic offset of almost $-1$\,dex. It would appear therefore to be perfectly acceptable to assume a canonical molecular abundance. If two species are available, and their abundance ratios are of interest, it should therefore be acceptable to fix one abundance and have as a free parameter the abundance ratio itself, thereby removing a free parameter. Further testing is needed however to see whether this result holds when the synthetic line ratios emerge from clouds with $K_{\rm vir}\neq 1$.

\begin{table*}
\centering
\caption{Deviations of Recovered Parameters from MCMC Models For the free and assumed $X_{\rm HCN}$.}\label{tab:FreeHCN}
\begin{tabular}{c c c c c c}\hline
& & \multicolumn{2}{c}{Free $X_{\rm HCN}$} & \multicolumn{2}{c}{Fixed $X_{\rm HCN}$} \\ 
& & Best & Mean & Best & Mean \\ \hline\hline
\multirow{4}{*}{\rotatebox[origin=c]{90}{Statistical}} & $T_{\rm k}$ & $-0.1\pm0.7(0.06)$ & $-0.1\pm0.6(0.06)$ & $0.0\pm0.7(0.08)$ & $0.0\pm0.6(0.08)$ \\
& $n_{{\rm H}_2}$ & $0.5\pm1.4(0.4)$ & $0.5\pm1.3(0.4)$ & $0.0\pm1.2(0.2)$ & $-0.1\pm1.2(0.2)$\\
& ${\rm d}v/{\rm d}r$ & $0.2\pm0.7(0.3)$ & $0.3\pm0.7(0.3)$ & $0.0\pm0.7(0.2)$ & $0.0\pm0.6(0.2)$ \\
& $X_{\rm HCN}$ & $-0.7\pm2.4(0.7)$ & $-1.2\pm2.1(0.7)$ & $0.2\pm2.3$ & $0.2\pm2.3$ \\ \hline
\multirow{4}{*}{\rotatebox[origin=c]{90}{Median}} & $T_{\rm k}$ & $-0.001\pm0.003$ & $-0.011\pm0.004$ & $0.044\pm0.005$ & $0.035\pm0.003$\\
& $n_{{\rm H}_2}$ & $0.32\pm0.05$ & $0.34\pm0.05$ & $-0.10\pm0.04$ & $-0.12\pm0.04$\\
& ${\rm d}v/{\rm d}r$ & $0.15\pm0.03$ & $0.18\pm0.03$ & $-0.05\pm0.03$ & $-0.05\pm0.02$ \\
& $X_{\rm HCN}$ & $-0.63\pm0.13$ & $-1.07\pm0.09$ & $0.06\pm0.14$ & $0.05\pm0.13$ \\ \hline 
\end{tabular}
\end{table*}

\subsection{Assumed [HCN]/[H$^{13}$CN]}

The results of the Assumed [HCN]/[H$^{13}$CN] model using one H$^{13}$CN line and with 10\% errors are shown in Figure \ref{fig:fix13c} and Table \ref{tab:fixH13CN}.

\begin{table}
\centering
\caption{Deviations of Recovered Parameters from MCMC Models for an Assumed [HCN]/[H$^{13}$CN].}\label{tab:fixH13CN}
\begin{tabular}{c c c c}\hline
& \multicolumn{3}{c}{Assumed [HCN]/[H$^{13}$CN]}  \\ 
& & Best & Mean \\ \hline\hline
\multirow{4}{*}{\rotatebox[origin=c]{90}{Statistical}} & $T_{\rm k}$ & $-0.1\pm0.7(0.03)$ & $-0.1\pm0.7(0.03)$ \\
& $n_{{\rm H}_2}$ & $-0.3\pm0.9(0.1)$ & $-0.3\pm0.9(0.1)$ \\
& ${\rm d}v/{\rm d}r$ & $-0.2\pm0.6(0.1)$ & $-0.2\pm0.6(0.1)$ \\
& $\frac{[{\rm HCN}]}{[{\rm H}^{13}{\rm CN}]}$ & $-0.3\pm0.6$ & $-0.3\pm0.6$ \\ \hline
\multirow{4}{*}{\rotatebox[origin=c]{90}{Median}} & $T_{\rm k}$ & $-0.042\pm0.014$ & $-0.032\pm0.009$ \\
& $n_{{\rm H}_2}$ & $-0.20\pm0.04$ & $-0.23\pm0.04$ \\
& ${\rm d}v/{\rm d}r$ & $-0.29\pm0.03$ & $-0.232\pm0.026$ \\
& $\frac{[{\rm HCN}]}{[{\rm H}^{13}{\rm CN}]}$ & $-0.282\pm0.028$ & $-0.281\pm0.029$ \\ \hline 
\end{tabular}
\end{table}

The striking result here is that the parameters are recovered much more poorly than in Figure \ref{fig:c12c13_1line}, where we ran the same model but without fixing [HCN]/[H$^{13}$CN] to an assumed value. Simultaneously, the MCMC trace reports much smaller errors. In other words, the MCMC fixes very tightly on the wrong point in $T_{\rm k} - n_{{\rm H}_2}$ space. I.e., we are actually better off adding [HCN]/[H$^{13}$CN] as a free parameter. In doing so we will not see as dramatic reductions in the parameter space as can be seen by fixing the [HCN]/[H$^{13}$CN] ratio, but we are also less likely to exclude the true parameters.

Fixing [HCN]/[H$^{13}$CN] to a canonical value causes three problems: {1)} due to the degeneracy between [HCN]/[H$^{13}$CN] and $T_{\rm k}$, fixing [HCN]/[H$^{13}$CN] artificially restricts $T_{\rm k}$, and if [HCN]/[H$^{13}$CN] is not close to the canonical value this then offsets $T_{\rm k}$. However, since $T_{\rm k}$ and $n_{{\rm H}_2}$ are also partially degenerate, this also offsets $n_{{\rm H}_2}$. Then, since the virial state of the gas is dependent upon $n_{{\rm H}_2}$ and d$v/$d$r$, d$v/$d$r$ is also affected (for a fixed $K_{\rm vir}$). {2)} as well as increased deviations, there is a systematic offset introduced into the physical parameters. {3)} an incorrectly assumed [HCN]/[H$^{13}$CN] can force $T_{\rm k}$, and hence $n_{{\rm H}_2}$, into unphysical regions where RADEX becomes unstable, leading to wildly incorrect parameter recovery. 

When using an MCMC there is a further issue, which is that the MCMC will dramatically understate the true uncertainty on the recovered parameters. Fixing [HCN]/[H$^{13}$CN] is a very strong prior, so that the MCMC will show only small oscillations about the mean parameters. This is precisely why $^{13}$C isotopologue line observations are so powerful: they are extremely good at breaking the degeneracies found in single species models. However, this means that if the wrong isotopologue abundance ratio is assumed then the recovered parameters will be forced to be incorrect. \emph{This effect is so large that you can actually be worse off assuming a canonical isotopologue abundance ratio than if you had no isotopologue lines at all.}

\subsection{$T_{\rm bg}$}\label{subsec:tbg_results}

\begin{figure*}
\centering
\includegraphics[width=\textwidth]{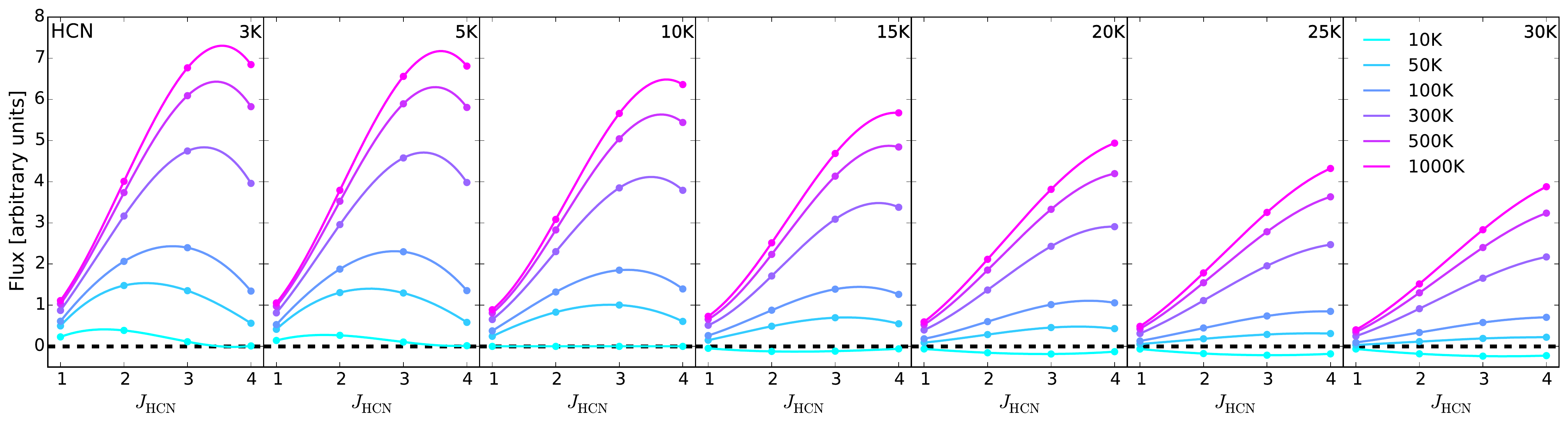}\\
\includegraphics[width=\textwidth]{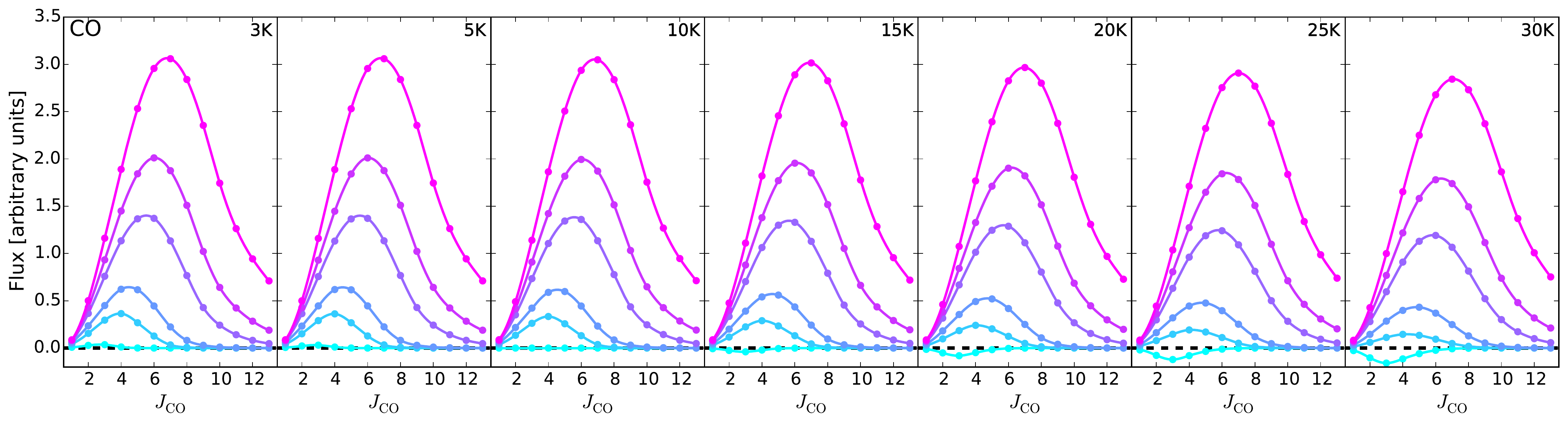}
\caption{The effect of increasing the background radiation temperature from 3\,K to 30\,K on the HCN SLED (top) and CO SLED (bottom) for a range of kinetic temperatures from 10\,K to 1000\,K (legend is common to both plots). $T_{\rm bg}$ increases from left to right as indicated, and $T_{\rm k}$ varies by colour, from sky blue (cold) to dark magenta (hot). Within each species all plots are to the same, arbitrary, flux scale.}\label{fig:tbg}
\end{figure*}

The HCN and CO SLEDs for ranges of $T_{\rm bg}$ and $T_{\rm k}$ are shown in Figure \ref{fig:tbg}. Each panel corresponds to the indicated $T_{\rm bg}$, increasing from left to right, and we plot the line flux in arbitrary units against the $J_u$ of the line transition. All panels share the same vertical scale. The colour corresponds to $T_{\rm k}$ from cold (10\,K, sky blue) to hot (1000\,K, dark magenta). While it is usually the case that $T_{\rm k}\geq T_{\rm dust} \geq T_{\rm CMB}$ we include regions where the kinetic temperature is lower than the incident blackbody radiation temperature for completeness. 

We note that these are the observable fluxes; i.e., the line flux seen in excess of the continuum level, which RADEX provides by default. This ``loss of contrast'' is the primary cause of the ``shrinking'' of the SLEDs as $T_{\rm bg}$ increases.

As expected, the SLEDs fall into absorption when their $T_{\rm k}$ is less than $T_{\rm bg}$. All of the fluxes fall as $T_{\rm bg}$ increases: this is due to the reduced contrast of the line emission against the hotter background radiation field \citep[e.g.,][]{Papadopoulos2010}. The CO SLEDs display fundamentally similar, if offset, behaviour to the HCN SLEDs.

This model is, as ever, a gross simplification of reality, where hot background radiation fields on 10\,pc scales are usually due to AGN or starburst heated dust which also emits significantly in the mid-IR, leading to optical pumping of molecular vibrational lines \citep{Aalto2007, GonzalezAlfonso2013, Aalto2015, Aalto2015b}. Nevertheless, the low $T_{\rm bg}$ SLEDs, where the background temperature of 10\,K is reached by the CMB at $z=2.3$ demonstrate that \emph{line ratios indicative of certain behaviours in the local universe are not necessarily applicable in high redshift galaxies}. Similarly, any investigation into correlations between line ratios and galaxy properties must take into account variation in the incident radiation field, as well as variation within individual galaxies. 

It is noteworthy that since RADEX presents the line brightness in excess of the background, if there is a hot background field, but it is completely obscured by the foreground gas and dust, then the background temperature should be added back onto the line brightness produced by RADEX. On the other hand, this very specific geometry is probably rare in nature, and may be hard to identify. Furthermore, in these cases it is highly likely that only the outer edges of the very dense clouds are visible, where the gas is not exposed to the full force of the intense background field, further complicating the analysis.

Line ratios are frequently invoked where there are few line observations of a source. In principle, they allow an exploration of the molecular gas conditions and there have been extensive efforts to develop diagnostic molecular line ratios, e.g., as a means of distinguishing between obscured AGN and obscured starbursts \citep{Kohno2003, Kohno2005, Meijerink2005, Meijerink2007, Imanishi2006, Krips2008}. More recently, \citet{Izumi2013} have attempted to develop diagnostic plots using the HCN$(4-3)$/HCO$^+(4-3)$ and HCN$(4-3)$/CS$(7-6)$ line ratios. However, these studies are frequently single dish observations of entire galaxies. There are therefore potential concerns that gas and dust temperature gradients in the galaxies may have significant effects on the integrated line ratios\footnote{On the other hand, in active galaxies the molecular line emission is dominated by emission from centre of the galaxy.}, especially if the ratios are dependent on $T_{\rm bg}$, which, in active galaxies in the local universe, is dominated by dust emission. If these effects are varying between galaxies then this may lead to either scatter in true relationships or anomalous relationships where none exist. Furthermore, extending these line ratios to higher redshift, where $T_{\rm CMB}$ is elevated, could present further problems. 

\begin{figure*}
\centering
\includegraphics[width=\textwidth]{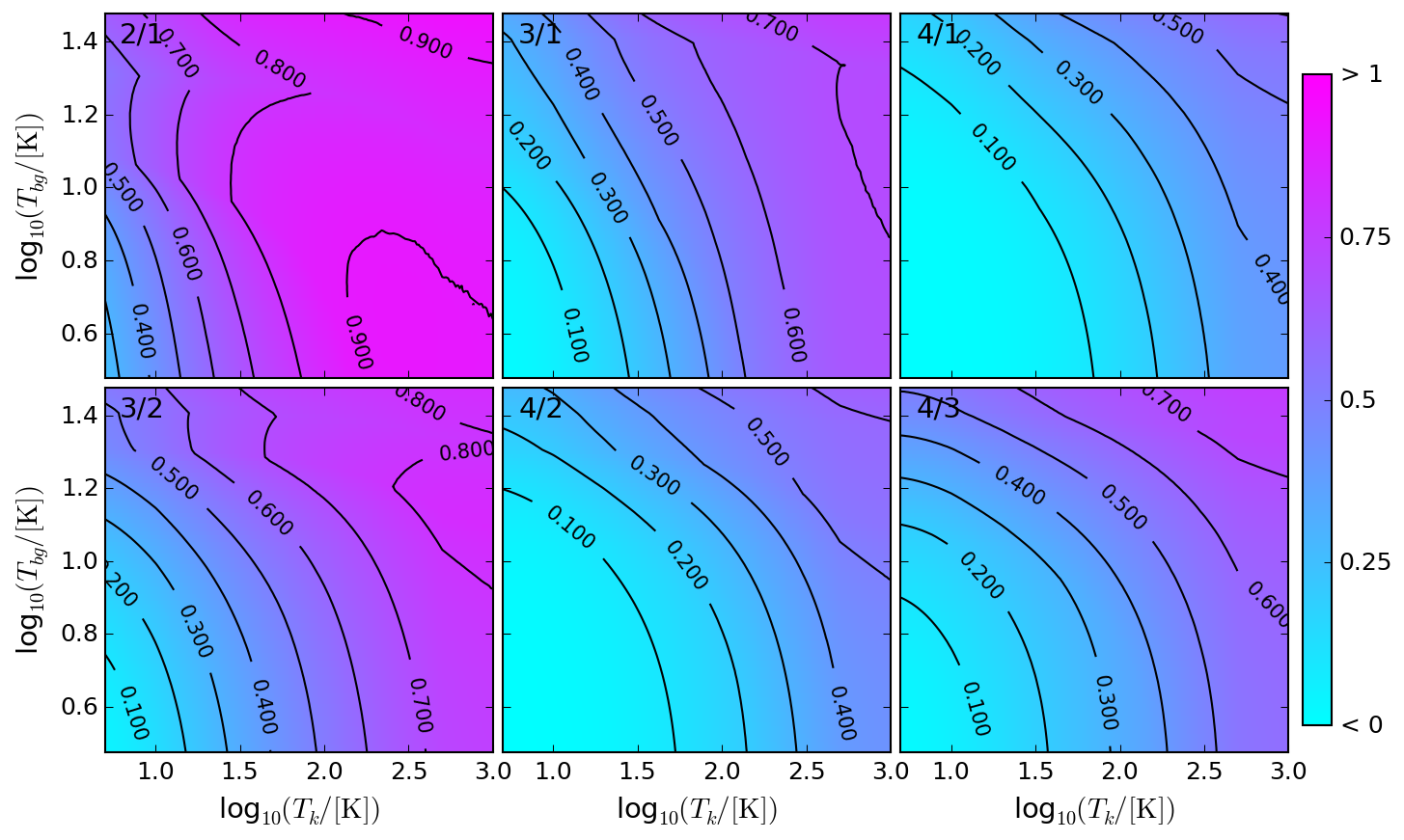}
\caption{The effect of increasing the background radiation temperature from 3\,K to 30\,K on the HCN line ratios for a range of kinetic temperatures from 10\,K to 1000\,K. The line ratio is given in the top right corner of each plot.}\label{fig:HCNtbgratios}
\end{figure*}

We present 2D plots of the HCN line ratios as functions of $T_{\rm k}$ and $T_{\rm bg}$ in Figure \ref{fig:HCNtbgratios}. There is a clearly visible degeneracy between $T_{\rm k}$ and $T_{\rm bg}$, especially in ratios with the $J=4-3$ line. While the low$-J$ line ratios do not change significantly, the HCN$\frac{(4-3)}{(3-2)}$ line ratio changes dramatically {between $T_{\rm bg}=3$\,K, $T_{\rm k}=10$\,K and $T_{\rm bg}=30$\,K, $T_{\rm k}=1000$\,K, rising rapidly.}

\bigskip

\begin{figure*}
\centering
\includegraphics[width=\textwidth]{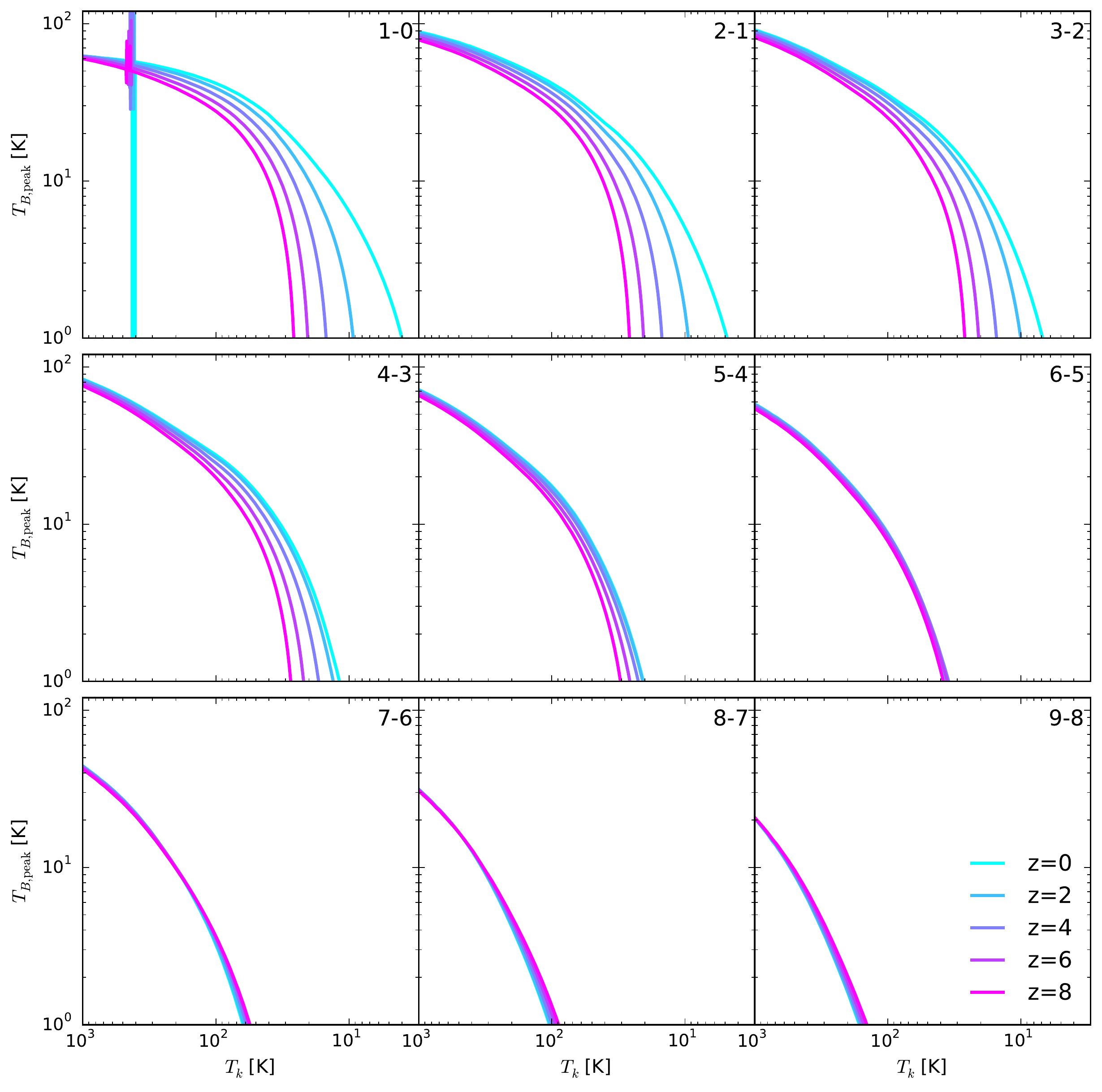}
\caption{The effect of $T_{\rm bg}$ on rest frame CO rotational line emission as a function of gas kinetic temperature. Note that the x-axis decreases {from left to right}. The rotational transition is given in the upper {right} corner of each subplot. The discontinuities in the $1-0$ plot are due to localised non-convergences in RADEX.}\label{fig:TbComp}
\end{figure*}

\begin{figure*}
\centering
\includegraphics[width=\textwidth]{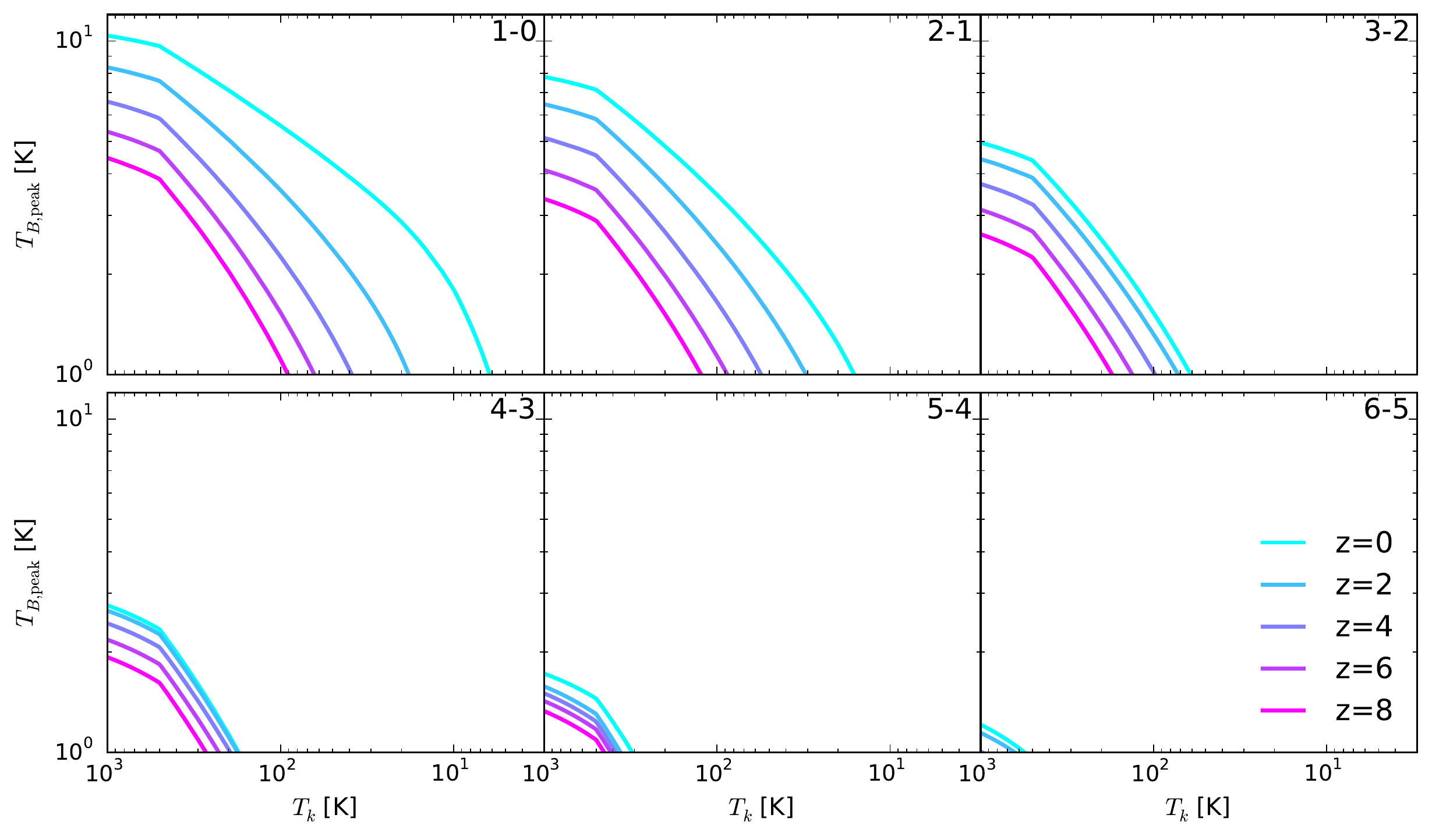}
\caption{The effect of $T_{\rm bg}$ on rest frame HCN rotational line emission as a function of gas kinetic temperature. Note that the x-axis decreases {from left to right}. The rotational transition is given in the upper {right} corner of each subplot. The kink at $T_{\rm k}>500$\,K is due to the lack of collisional rate data above this temperature for HCN.}\label{fig:TbCompHCN}
\end{figure*}

\noindent The results of our model exploring the effect of $T_{\rm bg}$ on $T_{\rm B, peak}$ as a function of $T_{\rm k}$ are shown in Figures \ref{fig:TbComp} and \ref{fig:TbCompHCN} where we plot the peak brightness temperature for the first 9 CO rotational transitions and first 6 HCN rotational transitions as functions of $T_{\rm k}$ for a range of redshifts. The models demonstrate how cooler molecular gas, such as the at outer edges of molecular reservoirs or in interclump regions, can become indistinguishable from the CMB. If galaxies at high $z$ possess kinetic temperature gradients {or harbour cold molecular gas environments}, {the aforementioned effect could lead to a bias in the gas morphologies inferred from low-$J$ CO lines}. This is not a new or surprising result; from a purely analytical approach this is to be expected since the total intensity is:

\begin{equation}
I_\nu^{J_u} \propto S_\nu^{J_u}\left(1-e^{-\tau_\nu^{J_u}}\right) + B_\nu(T_{\rm CMB})e^{-\tau_\nu^{J_u}},
\end{equation}

where $S$ is the source function, and therefore that the intensity in excess of the background field (here just the CMB) is:

\begin{align}
\Delta I_\nu^{J_u} &\propto S_\nu^{J_u}\left(1-e^{-\tau_\nu^{J_u}}\right) + B_\nu(T_{\rm CMB})\left(e^{-\tau_\nu^{J_u}}-1\right),\\
&= \left(1-e^{-\tau_\nu^{J_u}}\right)\left(S_\nu^{J_u} - B_\nu(T_{\rm CMB})\right),
\end{align}

so that when the gas is emitting approximately thermally, or $T_{\rm ex}\simeq T_{\rm CMB}$, then $S_\nu^{J_u}\simeq B_\nu(T_{\rm ex})$, and the line becomes indistinguishable from the CMB \citep[e.g.,][]{ToolsRadioAstronomy, daCunha2013}. This will introduce an additional bias towards detecting hotter, more excited gas as redshift increases.

The effect is more pronounced for HCN: this is due to the lines being naturally fainter than the CO lines (due to the much lower abundance of HCN cf CO), and so the fall in contrast against the CMB in much more pronounced. There is also a similar effect for dust, with the same associated implications on observability \citep{daCunha2013}.

\subsection{Toy Galaxy Models}

\begin{figure*}
\centering
\includegraphics[width=0.49\textwidth]{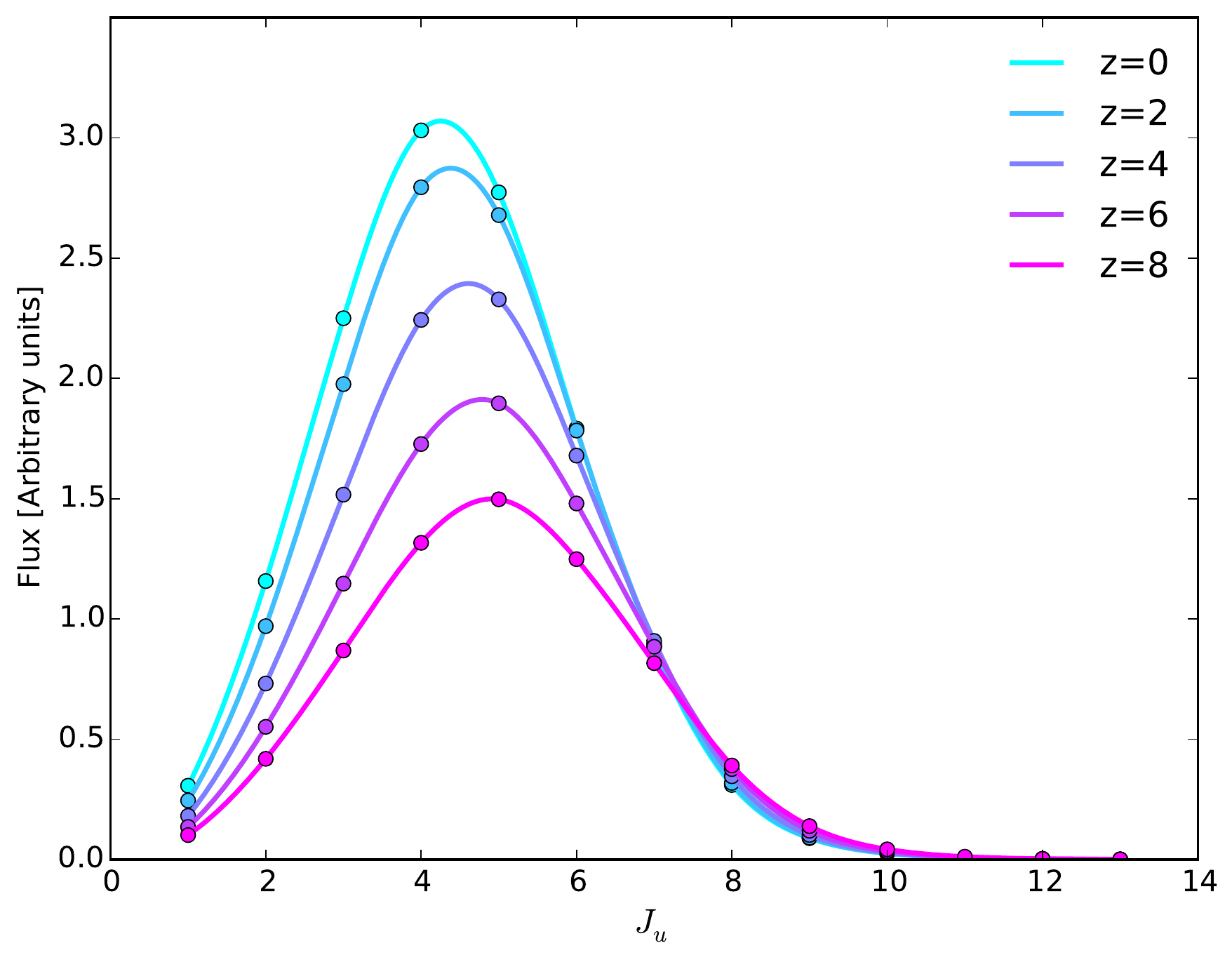}
\hfill
\includegraphics[width=0.49\textwidth]{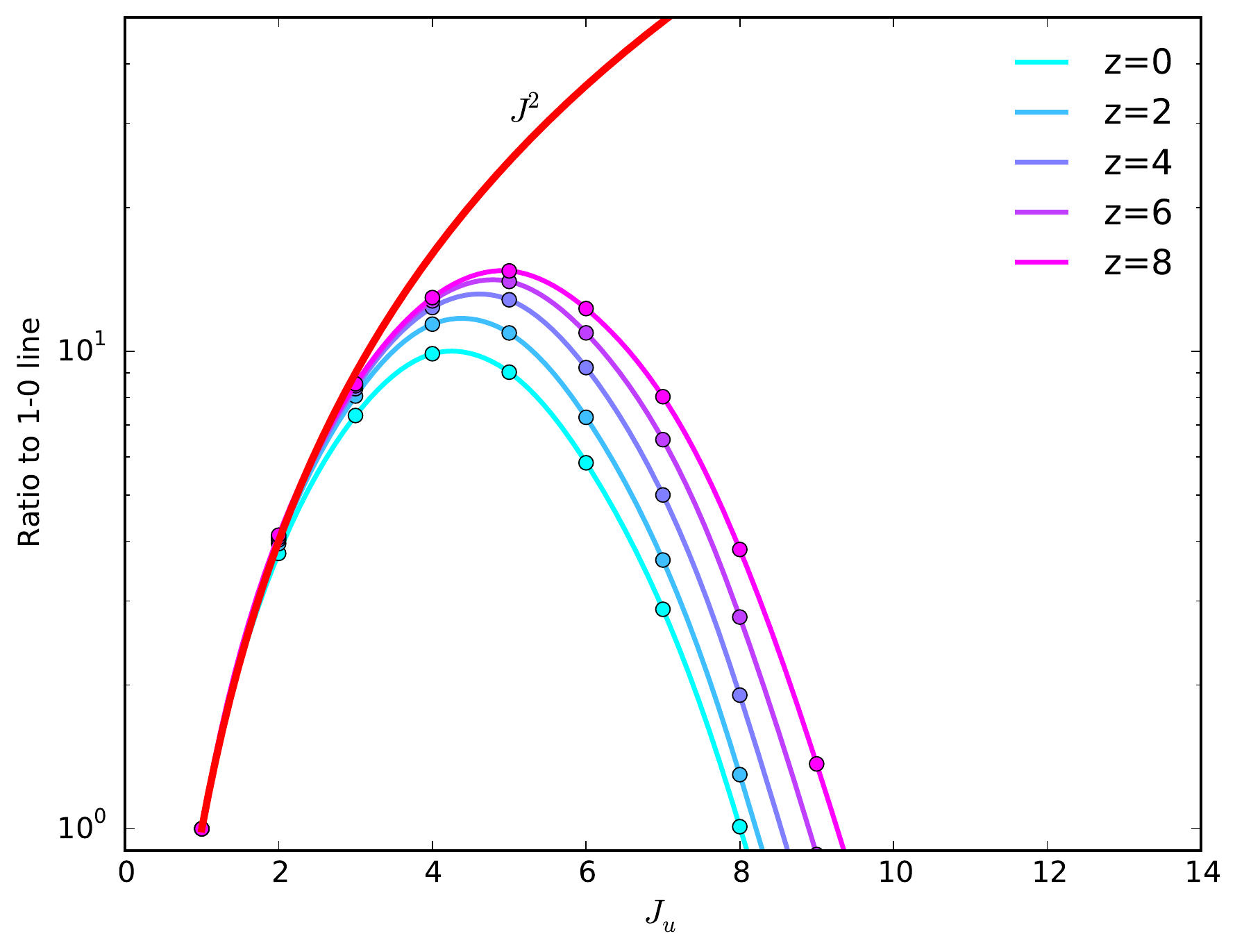}
\caption{Galaxy integrated CO SLEDs from $10^4$ clouds distributed exponentially with radius for five sample redshifts. The SLEDs are presented as arbitrary fluxes observed from the same distance (left), and as line intensity ratios to CO$(1-0)$ line (right), as in \citet{Narayanan2014}. The thick red line in the right hand plot is the thermalised $J^2$ limit. The trend of increasing excitation relative to the CO$(1-0)$ line in the high-$J$ lines as a function of redshift is due to the combination of reduced CO$(1-0)$ emission observable against the hotter CMB \emph{and} a small, real increase in the high-$J$ excitation due to the hotter CMB.}\label{fig:zcomp}
\end{figure*}

\begin{figure*}
\centering
\includegraphics[width=0.495\textwidth]{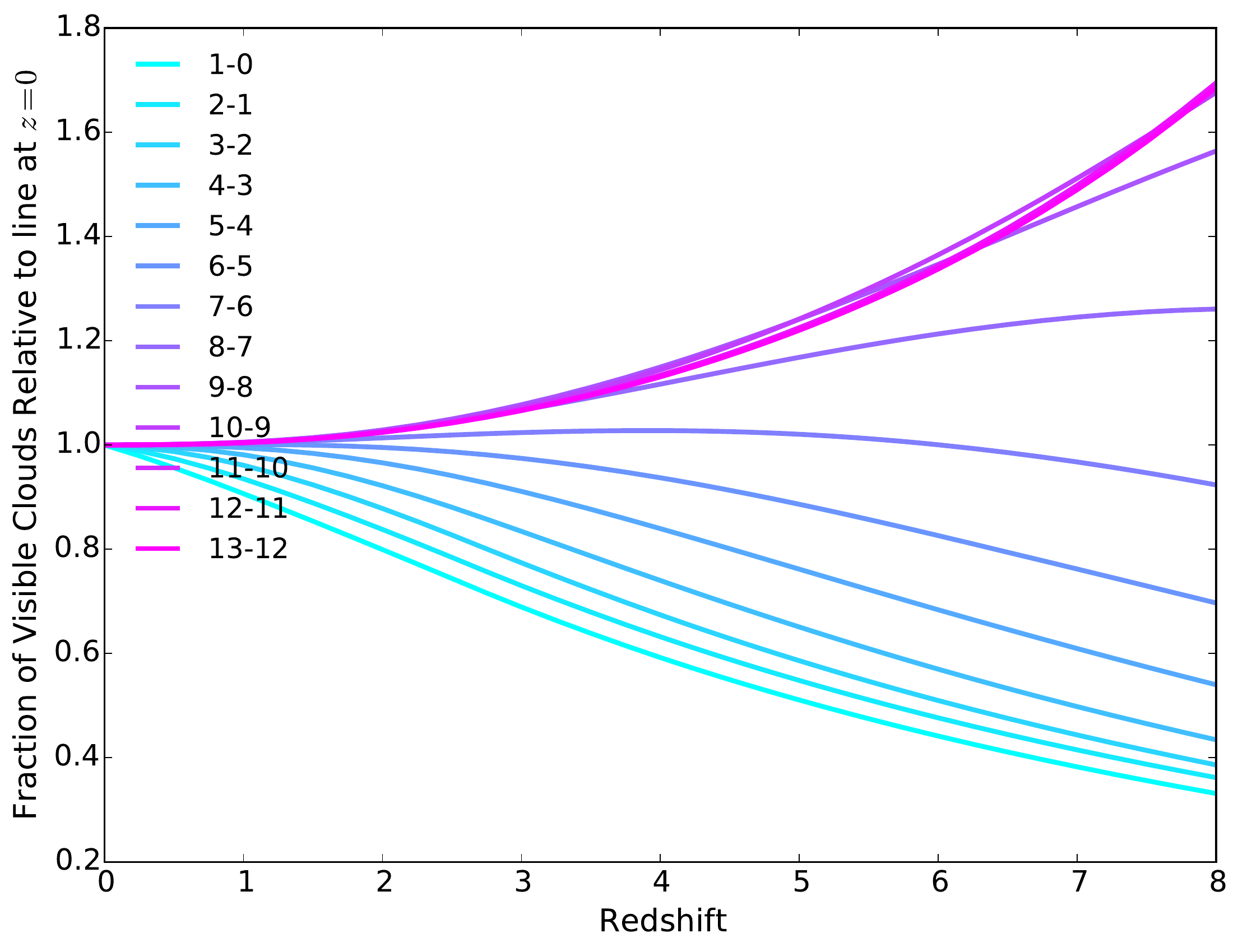}
\hfill
\includegraphics[width=0.495\textwidth]{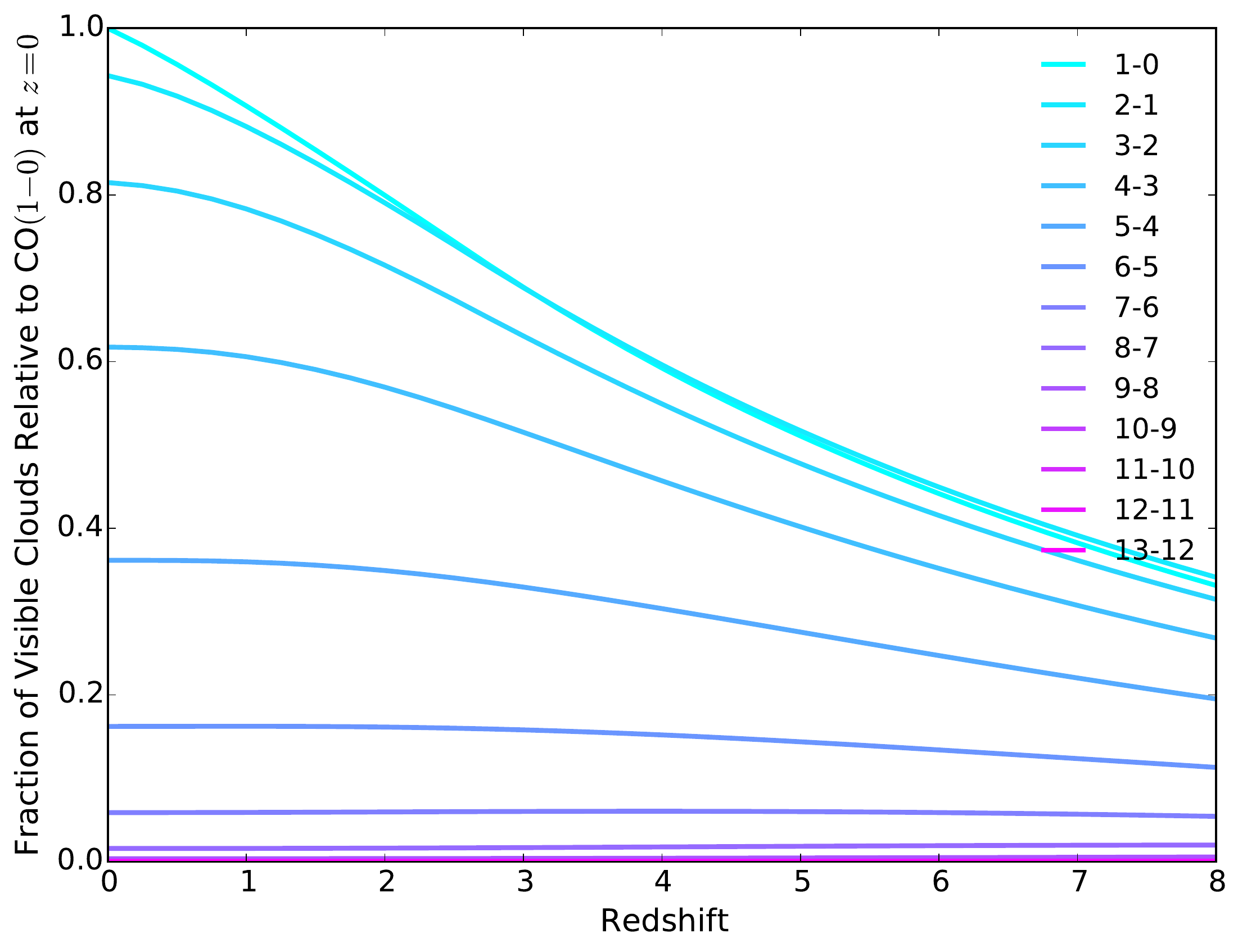}\\
\includegraphics[width=0.495\textwidth]{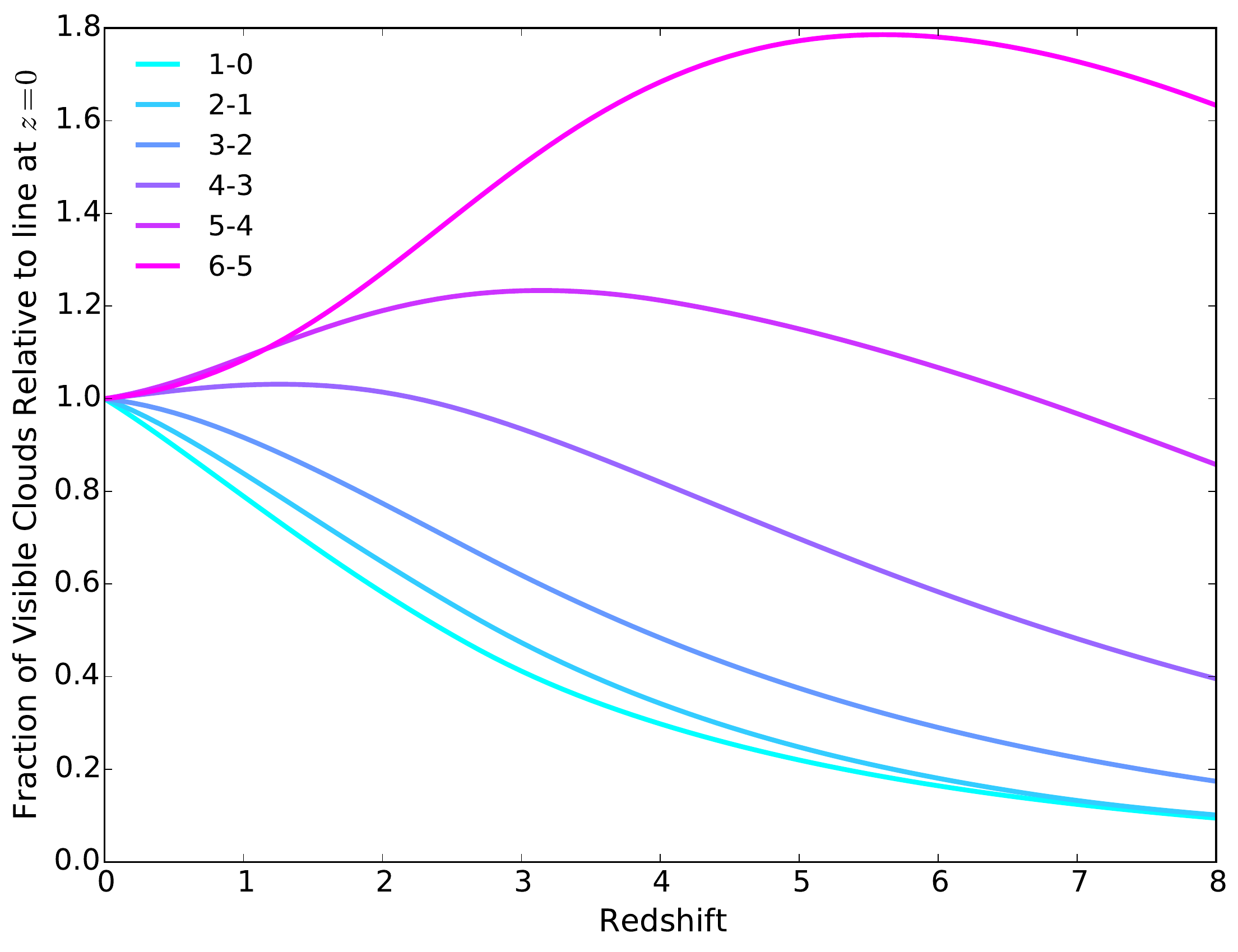}
\hfill
\includegraphics[width=0.495\textwidth]{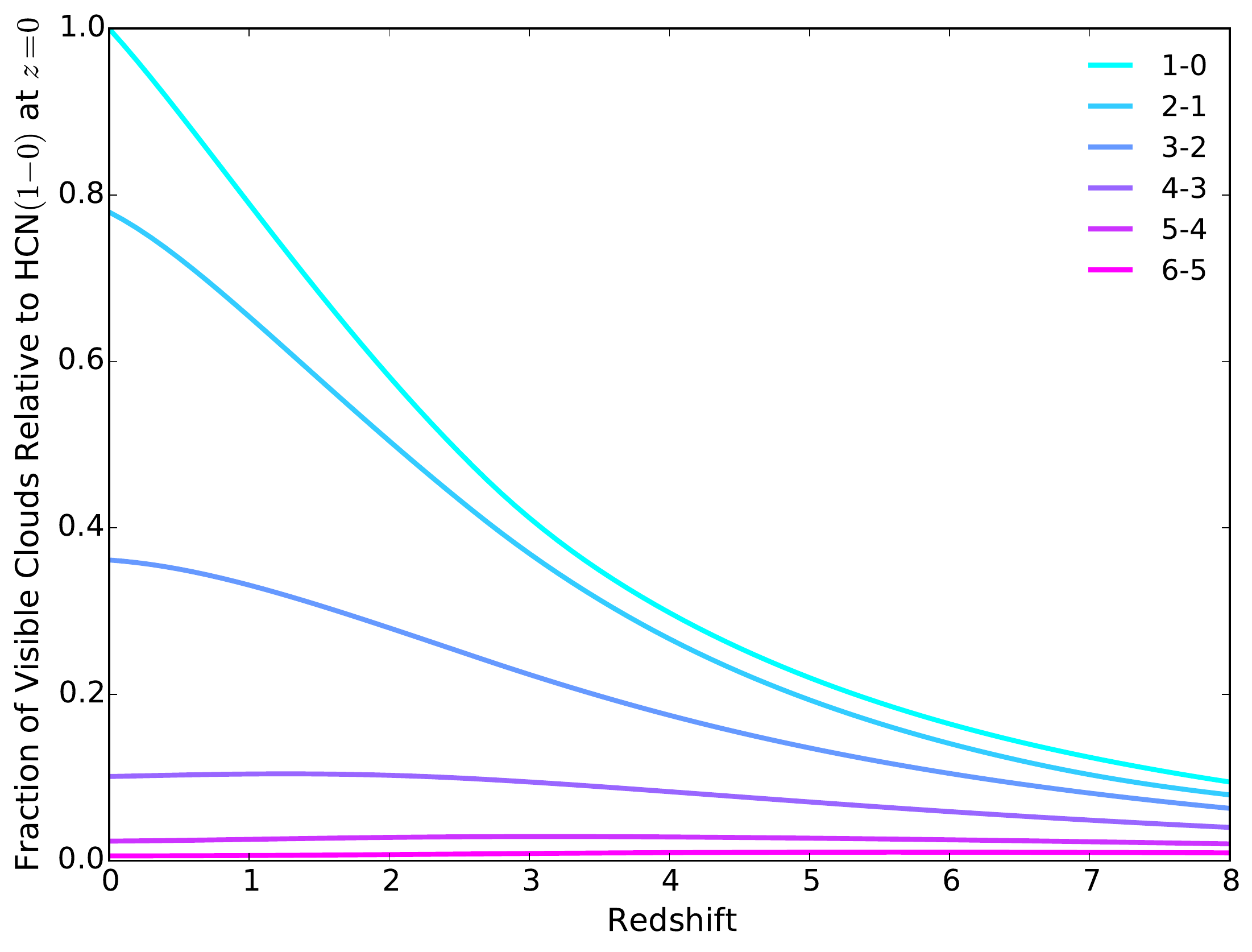}
\caption{Ratios of observed line luminosity relative to the line luminosity at $z=0$ (left) and relative to the $1-0$ line at $z=0$ (right) for the toy galaxy model, with CO (top) and HCN (bottom). Since the model has the same mass at all redshifts the CO$(1-0)$ track in these plots may be interpreted as one over the $\alpha_{\rm CO}$ correction factor necessary to account for the hotter CMB. Unlike the plots of \citet{daCunha2013}, which showed the effect of the CMB background on the intrinsic emission from the galaxy, these plots show the differences due to both decreased CMB contrast and the changes in line excitation due to the hotter CMB.}\label{fig:obsmass}
\end{figure*}

\begin{figure*}
\centering
\includegraphics[width=\textwidth]{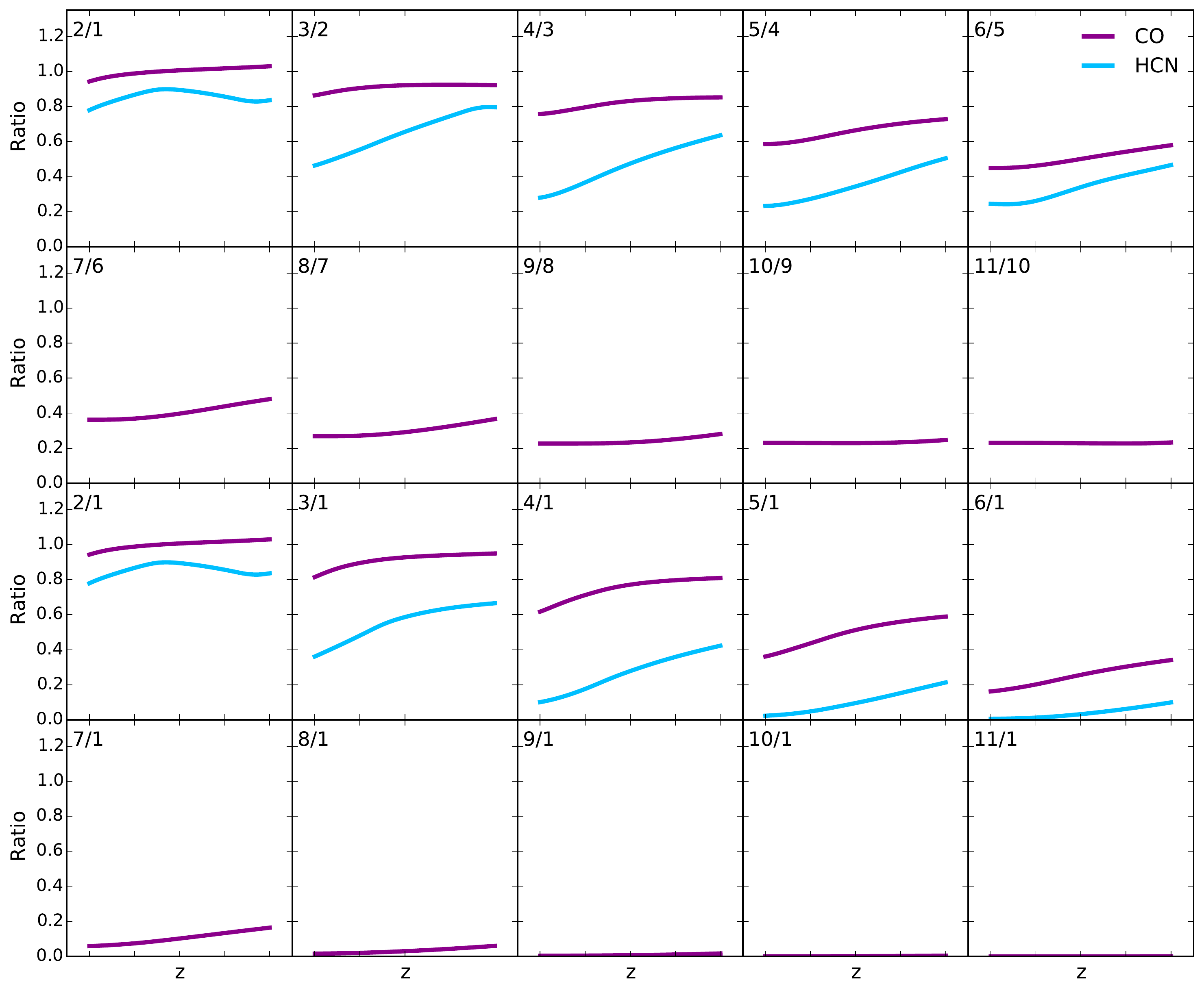}
\caption{Line brightness ratios for a selection of CO and HCN line ratios as functions of redshift, generated using the galaxy integrated model. The line ratios are indicated in the top left of each subplot. There is very little variation with redshift, except for in the $5-4/1-0$ and $4-3/1-0$ line ratios. These ratios are specific to the galaxy excitation model used here, and are not generally applicable. They are however indicative of the expected scale of the variation in the line ratios with redshift.}\label{fig:ratiosz}
\end{figure*}

The results of our toy model are shown in Figures \ref{fig:zcomp}, \ref{fig:obsmass} and \ref{fig:ratiosz}. As described in Section \ref{subsec:toys}, this model distributes $10^4$ clouds exponentially with radius across an artificial galaxy, with the clouds identical except for their $T_{\rm k}$, which is determined by their radius. The galaxy is then placed at a range of redshifts and the galaxy wide emission integrated. The SLED in Figure \ref{fig:zcomp}, left, shows how the flux, relative to the background, decreases with redshift, but also how the peak of the SLED shifts to higher $J$. These results are consistent with \citet{daCunha2013}.

The right hand SLED of Figure \ref{fig:zcomp} shows the same data, but with the SLED presented relative to the intensity of the $J=1-0$ line. These SLEDS are in turn consistent with \citet{Narayanan2014}. Note however that \citet{daCunha2013} and our Figure \ref{fig:zcomp}, left, show that the increase of the SLED with redshift, defined relative to the $J=1-0$ line, is primarily due to the much greater reduction in line contrast of the low$-J$ lines. While the increasing $T_{\rm bg}$ does indeed raise the high$-J$ lines this effect is small in comparison to the contrast reduction of the low$-J$ lines.

This reduction in the overall SLED is important, as it leads to a higher $\alpha_{\rm CO}$\footnote{$\alpha_{\rm CO} = M_{\rm mol}/L^\prime_{\rm CO(1-0)}$}{, e.g., equation 1 in \citet{Ivison2011} and equations 17 and 18 in \citet{Bolatto2013}}. Since the molecular gas mass of these models is identical at all redshifts, and since $L^\prime_{\rm CO(1-0)}$ ``de-redshifts'' emission (i.e., for two identical SLEDs at $z=0$ and $z=2$, excluding CMB effects, $L^\prime_{\rm CO(1-0)}$ will be identical) any difference in $L^\prime_{\rm CO(1-0)}$ in these models is solely due to changes in the CO$(1-0)$ line flux, which is solely due to the CMB. We show these effects in Figure \ref{fig:obsmass}, where we plot the line luminosities relative to the $z=0$ line luminosities (left) and the line luminosities relative to the $z=0$ CO$(1-0)$ line luminosity. For the adopted temperature profile and cloud distribution profile this effect is very large, with 80\% of the mass recovered at $z=2$ if a standard $\alpha_{\rm CO}$ is adopted, 60\% at $z=4$, and by $z=8$ only 35\% of the mass is recovered. 

We note that while  Figure \ref{fig:obsmass} left shows that the high-$J$ line luminosity increases significantly with redshift, Figure \ref{fig:obsmass} right shows that it still only traces a small fraction of the gas (for this toy model). The CO$(1-0)$ line at $z=0$ traces all of the molecular clouds, so the fraction of this luminosity recovered is directly comparable to the mass that would be estimated if a standard $z=0$ $\alpha_{\rm CO}$ factor was used. The HCN emission only traces the dense molecular gas, so the fall in the HCN integrated emission is representative of the dense gas that would be missed, not the total molecular mass.

\bigskip

\noindent In Figure \ref{fig:ratiosz} we present the line brightness temperature CO and HCN line ratios for our toy model galaxy. We present a selection of line ratios only. It is clear that there is very little variation in the galaxy integrated line ratios with redshift, except for in the $\frac{5-4}{1-0}$ and $\frac{4-3}{1-0}$ line ratios, although in all cases there is greater variation in the HCN ratios than the the CO ratios. This is, as in the $T_{\rm bg}$ investigation above, due to HCN simply being fainter than CO, so is more strongly affected by the brightening of the CMB. However, these ratios are specific to the galaxy excitation model used here, and are not generally applicable: there is likely to be a set of ratios which vary significantly, but the particular ratios will depend upon the peaks of the galaxy SLEDs, which in turn depends upon excitation conditions. The excitation conditions used here are rather weak, especially compared to the very active star forming models of \citet{Narayanan2014}. In general however, the galaxy integrated line ratios appear to be rather robust to changes in redshift, unlike the individual line of sight HCN brightness temperature ratios in Section \ref{subsec:tbg_results}.

\section{Discussion}\label{sec:discussion}

\subsection{Are LVG Model Results Reliable?}

LVG models are quick and easy to run, and have been used to a varying extent in Galactic, local universe and high redshift studies of molecular gas. One of most common, and indeed most reliable uses of the models is to generate grids of results, then overlaying either $\chi^2$ or likelihood contours \citep[e.g.,][]{GarciaBurillio2000,Rangwala2011}. This graphically illustrates the reasonable parameter ranges, and does not necessarily claim specific values. However, even these methods are susceptible to assumptions, which can bias the results in unexpected and unusual ways, especially if canonical abundances or abundance ratios are assumed. For example, assuming an [HCN]/[H$^{13}$CN] ratio of 60 actually imposes strong constraints on the possible $T_{\rm k}$ range. If the true [HCN]/[H$^{13}$CN] ratio is closer to 300 (such as in NGC\,6240: \citealp{Papadopoulos2014,Tunnard2015b}), then the assumptions can throw off $T_{\rm k}$ by over a dex. Furthermore, when multiple lines and species are available these approaches can be restrictive: they cannot be used to explore higher dimensional parameter spaces.

One great reassurance of the tests presented here is that there is no significant systemic offset in the model deviations, with most trials returning centrally peaked deviation distributions. In almost all models there was a small set of extreme outliers, almost always with an underestimate of $T_{\rm k}$ and concomitant overestimate of $n_{{\rm H}_2}$. These outliers are greatly exacerbated by including regions with a thermal pressure $>10^8$\,K\,cm$^{-3}$. This regime is so extreme that its exclusion is unlikely to remove many realistic regions with significant emission, especially when integrated across a galaxy.

The concern arises when model parameters are interpreted as hard and fast values, or as being known better than within a factor of 3. Even in the most optimistic scenarios and with zero observational uncertainties LVG models cannot be expected to provide answers more accurately than $\pm0.2$\,dex. The real uncertainties on the results are most likely much higher, presenting real and significant concerns with any possible precision uses of LVG models. 

\bigskip

\noindent A topic we have not mentioned until now is that of the collisional rate constants. LVG models require these (temperature dependent) rates which are primarily determined in laboratory experiments. In the models presented here we should be largely insensitive to any errors in these rates, as we are using the same rates for generating both the original lines and comparison lines (whether they be in grids or MCMC). In real world observations however, these errors could be significant, especially for the less well studied molecules (such as HNC, a chemically interesting isomer of HCN). Furthermore, RADEX does not extrapolate the rate constants beyond the provided data, keeping them constant instead\footnote{See \url{http://home.strw.leidenuniv.nl/~moldata/radex.html}}. We are not suggesting that this is the wrong approach; indeed any extrapolation of rate constants must be performed with utmost caution. We are simply noting that this should be appreciated as a limitation of LVG, and indeed any molecular line radiative transfer models, including LIME. Nevertheless, we have no means of quantifying these potential errors with our models, and as such consider the issue beyond the scope of this paper, leaving it as a further reason to consider the model precisions presented here as best cases only.

\bigskip

\noindent No discussion of the reliability of LVG models would be complete without mentioning chemistry. The LVG framework does not include chemistry; chemistry can be added to LVG models ante hoc, post hoc or iteratively, but in all cases this is just altering the molecular abundances. In other words, LVG codes, e.g., RADEX, can find excellent fits for completely unrealistic physical conditions and molecular abundances. 

To some extent unrealistic physical conditions may be excluded by the use of appropriate priors. Similarly, a sceptical interpretation of the LVG results is essential, preferably including an examination of the appropriateness of the derived molecular abundance ratios for the physical conditions. A potentially very powerful technique for this would be post hoc chemical modelling or comparison with lookup tables, to verify the results are self consistent, somewhat in the fashion of \citet{Viti2014}. Unlike running chemistry for every step in the MCMC, this method would not add significantly to the run time  while providing significantly more physical insight and potentially highlighting interesting results. 

\bigskip

\noindent We also point to the risks of modelling emission from entire galaxies as single clouds. Not only is this physically unmotivated, it produces unrealistic column densities and masses. Dynamically motivated and self consistent methods exist which are also physically meaningful, and although they may require prior assumptions these assumptions are in all cases more reasonable than the tacit assumptions implicit in the single cloud alternative \citep[see appendices in][]{Papadopoulos2012,Tunnard2015b}.

\subsection{An Aside on $K_{\rm vir}$}

There are some subtleties associated with assuming a virial state for the modelled gas. Since, for a uniform density cloud, the mass (in kg) is given by:

\begin{equation}
M_{\rm cloud} = \frac{4\pi}{3}\left(\frac{\zeta\, \Delta v}{2{\rm d}v/{\rm d}r}\right)^3 n_{{\rm H}_2}\,\mu_{{\rm H}_2},
\end{equation}
where $\mu_{{\rm H}_2}$ is the helium corrected mass per H$_2$ molecule, and $\zeta=3.08\times10^{18}$\,cm\,pc$^{-1}$, or equivalently that:
\begin{equation}
r_{\rm cloud} = \frac{\zeta\, \Delta v}{2{\rm d}v/{\rm d}r}.
\end{equation}
The issue is that if a virial state is assumed, then for an observed line width the size of the cloud is fixed. Therefore, if the assumed virial state of the cloud is changed (through d$v/$d$r$), and all other variables held constant, then the size of the cloud has also been changed. This subtlety is important both for resolved and extragalactic observations: if the size of the cloud is known from observations then this should be fixed in the modelling. If the size of the cloud is unknown, then there is a degeneracy between the size of the cloud and multiple clouds distributed in velocity space across the galaxy. 

\begin{figure*}
\centering
\includegraphics[width=0.495\textwidth]{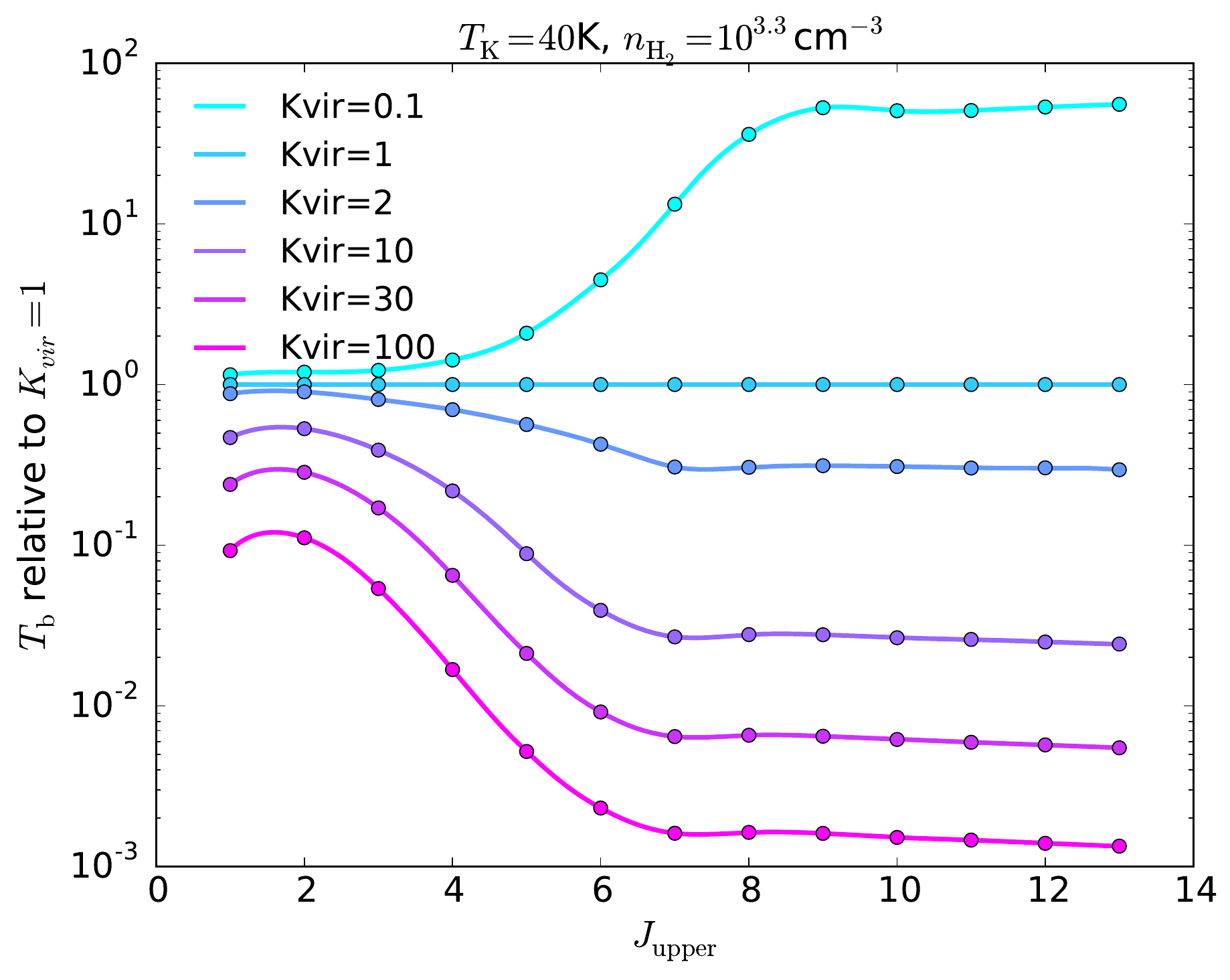}
\hfill
\includegraphics[width=0.495\textwidth]{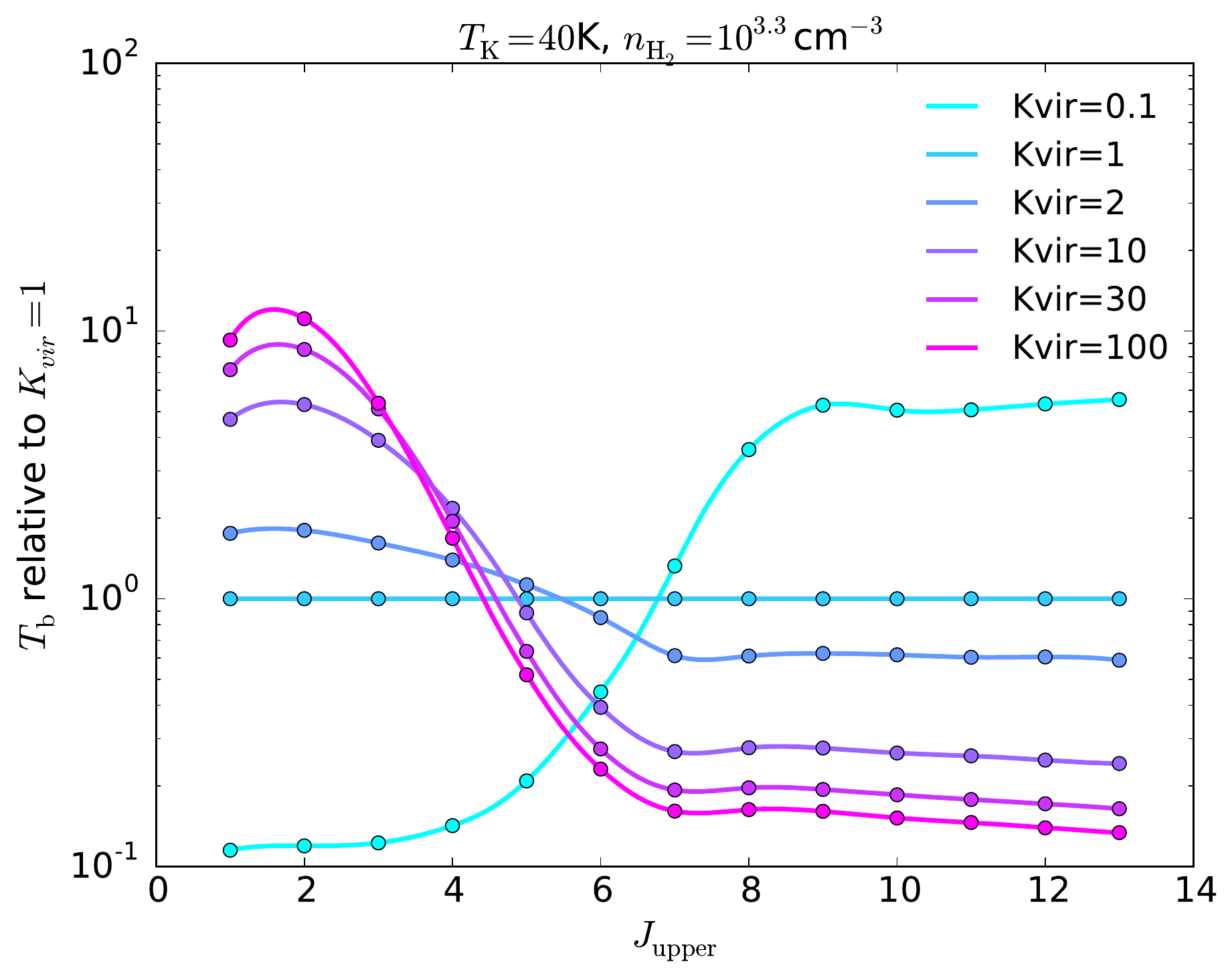}
\caption{The effect of varying $K_{\rm vir}$ without constraining cloud size (left) and constraining the cloud size to be 1\,pc (right). The peak brightness temperatures are presented relative to the $K_{\rm vir}=1$ case.\\}\label{fig:kvir}
\end{figure*}

Further discussion of this topic can be found in \citet{Papadopoulos2012} and \citet{Tunnard2015b}. Here, we simply present a comparison of the peak brightness temperature from RADEX models where $K_{\rm vir}$ is varied, first without any constraint on the cloud size, and then with the cloud size constrained to be 1\,pc, with the results shown in Figure \ref{fig:kvir}. In the unconstrained case there is less emission in all lines as $K_{\rm vir}$ increases: this is due to the cloud shrinking, and the column density of CO falling. In the $r=1$\,pc case however there is a pivot around the $J_u\simeq5-6$ lines, with emission increasing with $K_{\rm vir}$ at lower $J$ and decreasing at higher $J$. Here, the column density is the same but the lines become less optically thick as $K_{\rm vir}$ increases, so more of the column can be seen, and emission increases.

The key point here is that there are strong variations in the line emission as well as the relative line emission with variations of $K_{\rm vir}$ and dependent upon the underlying assumptions. In situations where the cloud size is known this can be particularly important.

\subsection{How important is $T_{\rm bg}$?}

We have presented several models of the potential effects of the background radiation field on LVG model results. These models do not, and cannot, account for dust within the emitting gas, as this is not a feature included in RADEX. They can however model gas illuminated by a separate hot dust background, and the effects of the CMB at higher redshift.

As we have noted, it is also important to recognise that RADEX presents line results as line brightnesses in excess of the background field. When dealing with the CMB this is ideal, however, when dealing with a hot dust source that is completely obscured by cooler, intervening dust this is misleading. In these cases it may be more appropriate for the applied background field to be re-added to the line brightnesses. On the other hand, due to the high column depths and line optical depths of CONs we are only able to observe the cooler, shielded outer layers in the major isotopologue lines \citep{Aalto2015b} at which point the background dust has cooled significantly and a high $T_{\rm bg}$ is not actually incident on the observable regions.

We have demonstrated the well known degeneracy between $T_{\rm k}$ and $T_{\rm bg}$, and demonstrated the effects of a hot dust background on HCN and CO SLEDs (Figure \ref{fig:tbg}). These effects can be huge, including the effects on the line ratios. However, these conditions are likely to be significant in a minority of objects, and more comprehensive modelling of dust included in the gas phase is beyond the scope of this paper. \citet{Papadopoulos2010} argued that if the gas and dust are in thermal equilibrium, then any heating of the dust will increase the line and continuum brightness equally, thereby reducing the line/continuum contrast. While essential when analysing objects such as Arp\,220, this is unlikely to be a concern in most galaxies.

\bigskip

\noindent At $z=4$ both CO$(4-3)$ and CO$(5-4)$ lie in ALMA Band 3 while the lower $J$ lines, observed by \citet{Carilli2010, Carilli2011, Hodge2012}, require the JVLA. These studies have revealed a clumpy medium: precisely what is expected if the real conditions are GMC-like clumps embedded in an extended diffuse medium. The low excitation diffuse medium becomes undetectable against the hotter CMB and only the high density clumps are seen \citep[see also][]{Casey2015}. The effects on planned blind surveys of CO with the ngVLA and SKA are more complex, but will risk introducing a significant redshift bias on both the low-$J$ CO luminosity function and the H$_2$ mass fractions, as well as the cosmological $\Omega ({\rm H}_2)$ \citep{Casey2015,Leroy2015}. How significant this effect will be is not clear, as it depends heavily on the kinetic temperature of the molecular gas at high $z$. If large fractions of the molecular reservoirs are $\lesssim 20$\,K then the effects are likely to be very significant as early as $z\simeq3-4$. However, if high $z$ molecular reservoirs are mostly $\geq30$\,K then we do not expect any significant effect even out to $z=8$.

The change in $\alpha_{\rm CO}$ with redshift highlights the importance of dynamical mass estimates for high $z$ line observations \citep{Carilli2010, Carilli2011, Hodge2012}, and presents new challenges for high $z$ mergers, where dynamical mass estimates may no longer be meaningful. Our model, while very simple, demonstrates that even at $z=2$ there are risks of significantly underestimating the molecular masses of galaxies due to the loss of contrast against the CMB, including the potential invisibility of the coldest, yet perhaps in some cases also the densest, regions of the galaxies.

The line ratios most affected by the higher CMB are those including the low$-J$ lines, so that higher $J$ line ratios should remain relatively robust even at redshifts as high as $z=8$, depending upon the galaxy excitation.


\section{Conclusions}\label{sec:conclusions}

We have presented tests of the ability of grid and MCMC methods to recover the true input parameters of LVG models using line ratios generated by the same LVG model. We have shown that without any additional errors there is a factor of $2-3$ uncertainty in most derived parameters. When a 10\% absolute flux calibration error is applied to the synthetic lines the parameter uncertainty can expand to a factor of 10, even when making very generous assumptions.

We have shown that the MCMC approach, used in \citet{Tunnard2015b} and similar to that of \citet{Kamenetzky2014}, produces results comparable to or better than the grid method, while also being easily extended to accommodate additional parameters, molecules and gas phases. In a future work we intend to examine the potential biases involved when observing an entire galaxy, reducing the continuum of conditions to three phase models.

Most importantly, we have demonstrated that assuming an isotopologue abundance ratio will lead to less accurate results than modelling it as a free parameter. Incorrect assumptions with regards to the isotopologue abundance ratio can have disastrous consequences for the accuracy of recovered $T_{\rm k}$ and $n_{{\rm H}_2}$ (and of course [HCN]/[H$^{13}$CN]). \emph{This effect is so large that you can actually be worse off assuming a canonical isotopologue abundance ratio than if you had no isotopologue lines at all.}

\bigskip

\noindent By examining the effect of the background radiation temperature on molecular line ratios we have demonstrated the potential for significant effects on line ratios due to hot dust backgrounds and the high $z$ CMB. Using toy models to explore the effects of the CMB on CO SLEDs with increasing redshift we demonstrate that for galaxies with molecular gas components with $T_{\rm k, z=0}\leq20$\,K {both morphologies and masses inferred from low$-J$ CO lines are likely to be affected when} $z\geq5.5$, while if there are molecular gas components with $T_{\rm k, z=0}\leq10$\,K there are large effects {as early as} $z\geq2$. The specific values of the biases are dependent upon the galactic $T_{\rm k}$ profile. Nevertheless, we find that galaxy integrated CO line ratios are likely to be fairly robust even out to $z=8$, whereas the line of sight line brightness ratios vary very strongly \citep[see also][]{daCunha2013}. Furthermore, it is imperative when modelling molecular rotational lines that the redshift is taken into account through the inclusion of the CMB. When modelling line ratios it is still important to include the CMB, but less so than when modelling the lines themselves.

The potentially very strong effect of the CMB on the recovered line flux has worrying implications for the application of $\alpha_{\rm CO}$ to high redshift studies, adding another complication to the already difficult mix of kinematic state, metallicity and chemistry.

\acknowledgements 
{\small This research is supported by an STFC PhD studentship. TRG acknowledges support from an STFC Advanced Fellowship. We extend special thanks to S.\ Viti for her knowledgable advice and guidance. We thank the anonymous referee for their comments and advice.}

\FloatBarrier
\bibliography{LVGtesting}
\FloatBarrier
\appendix
\FloatBarrier
\section{Additional Plots}

Here we include plots described throughout the paper which have been moved to the Appendix to improve the readability of the main paper.

\begin{figure*}[h]
\centering
\includegraphics[width=\textwidth, height=0.4\textheight]{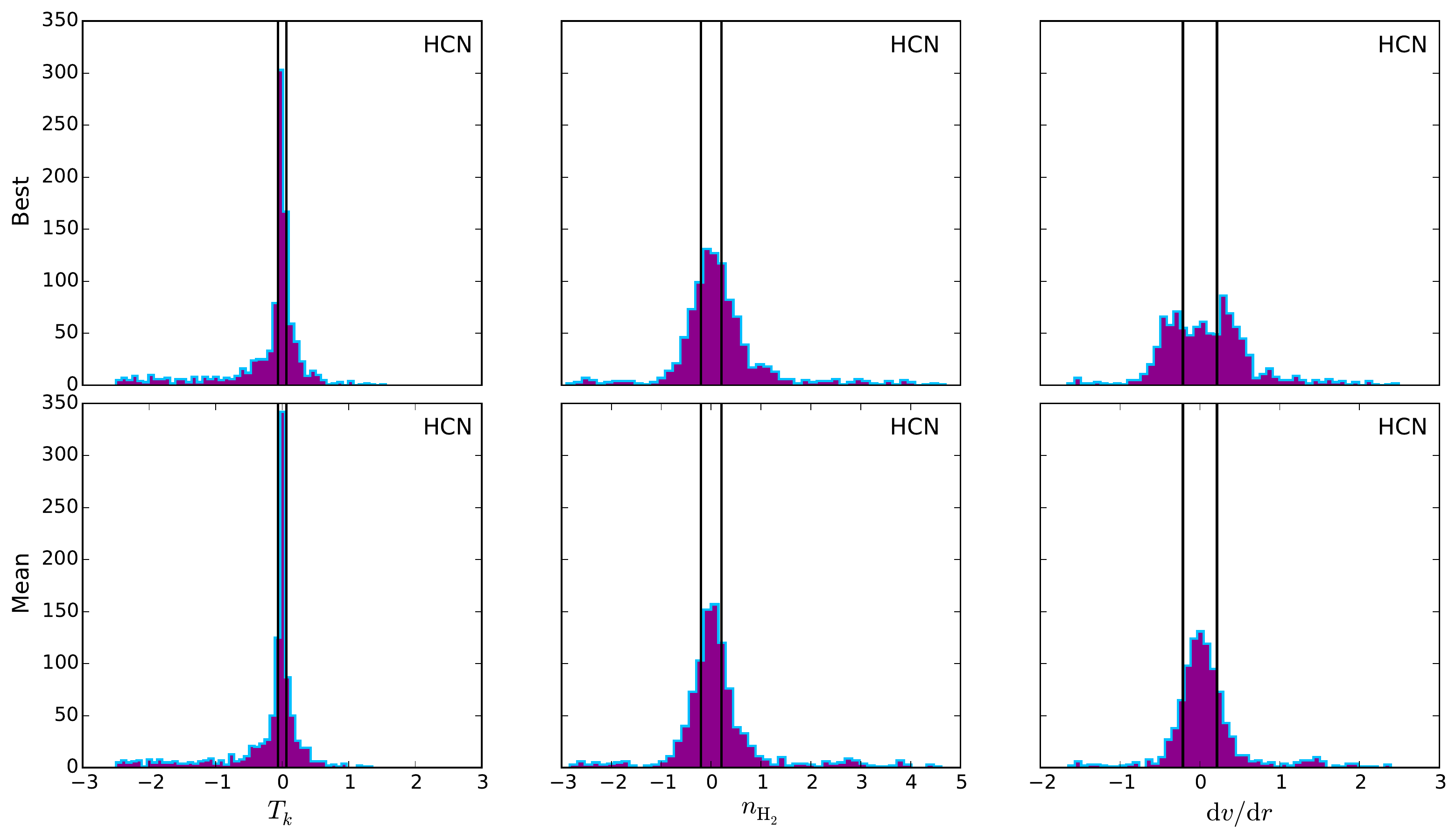}\\
\includegraphics[width=\textwidth, height=0.4\textheight]{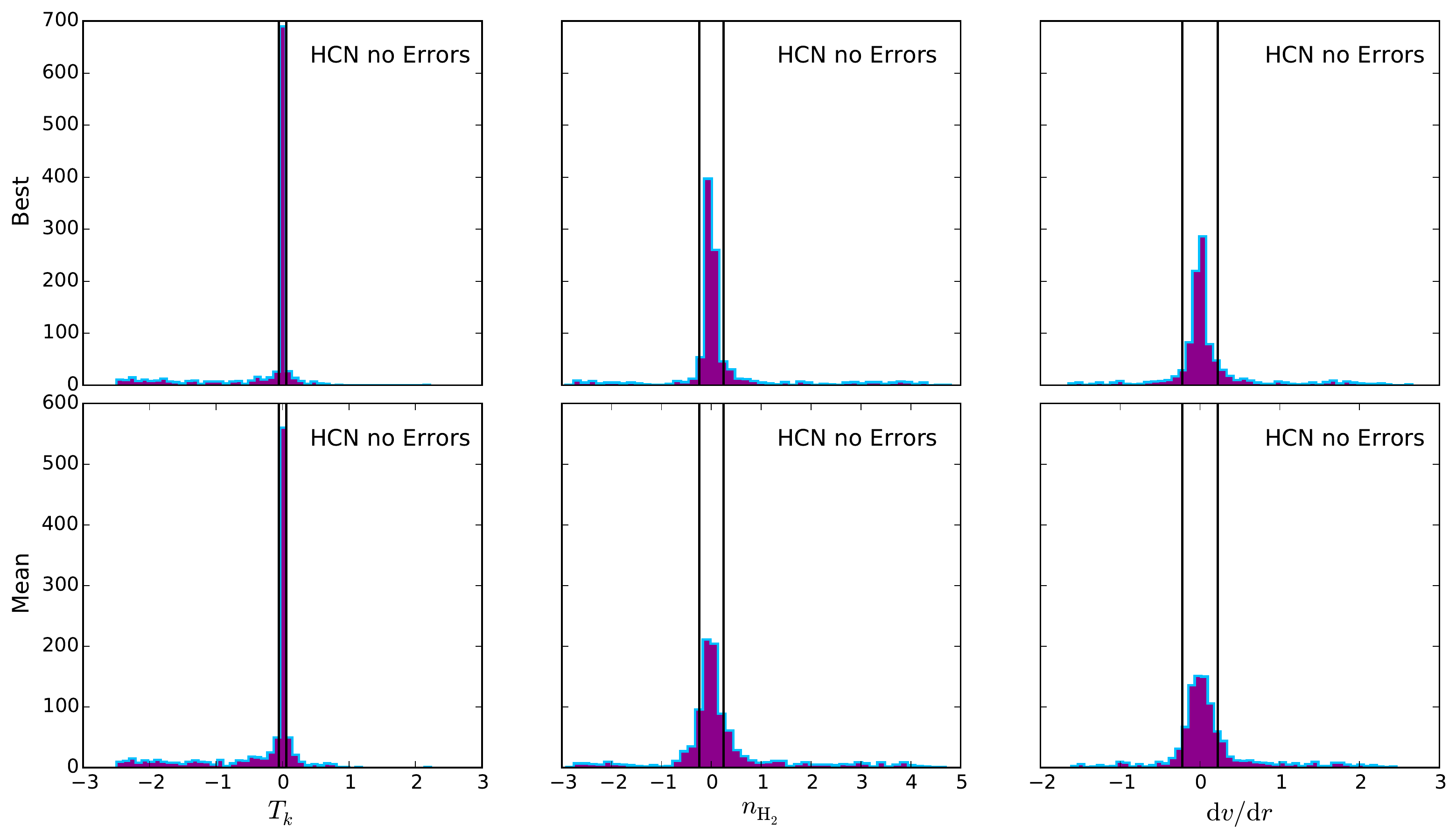}
\caption{The logarithm of the ratio of the recovered to true parameters with (upper panel) and without (lower panel) $10\%$ errors for HCN, drawn from a pressure restricted parameter range and recovered with an MCMC. The best fit (top row) and mean (bottom row) recovered parameters are reported.}\label{fig:MCMChcn}
\end{figure*}

\begin{figure*}[h]
\centering
\includegraphics[width=\textwidth, height=0.4\textheight]{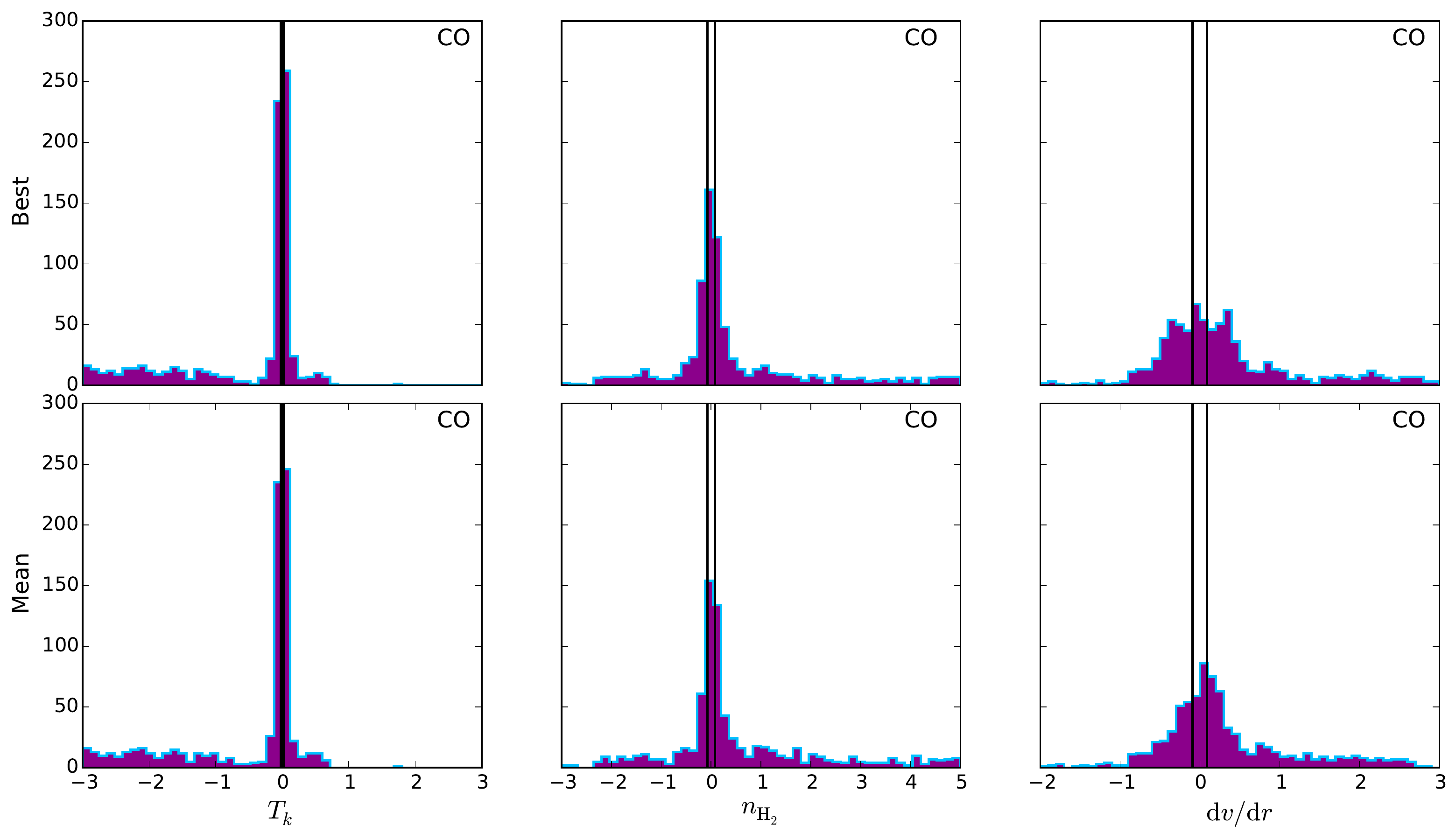}\\
\includegraphics[width=\textwidth, height=0.4\textheight]{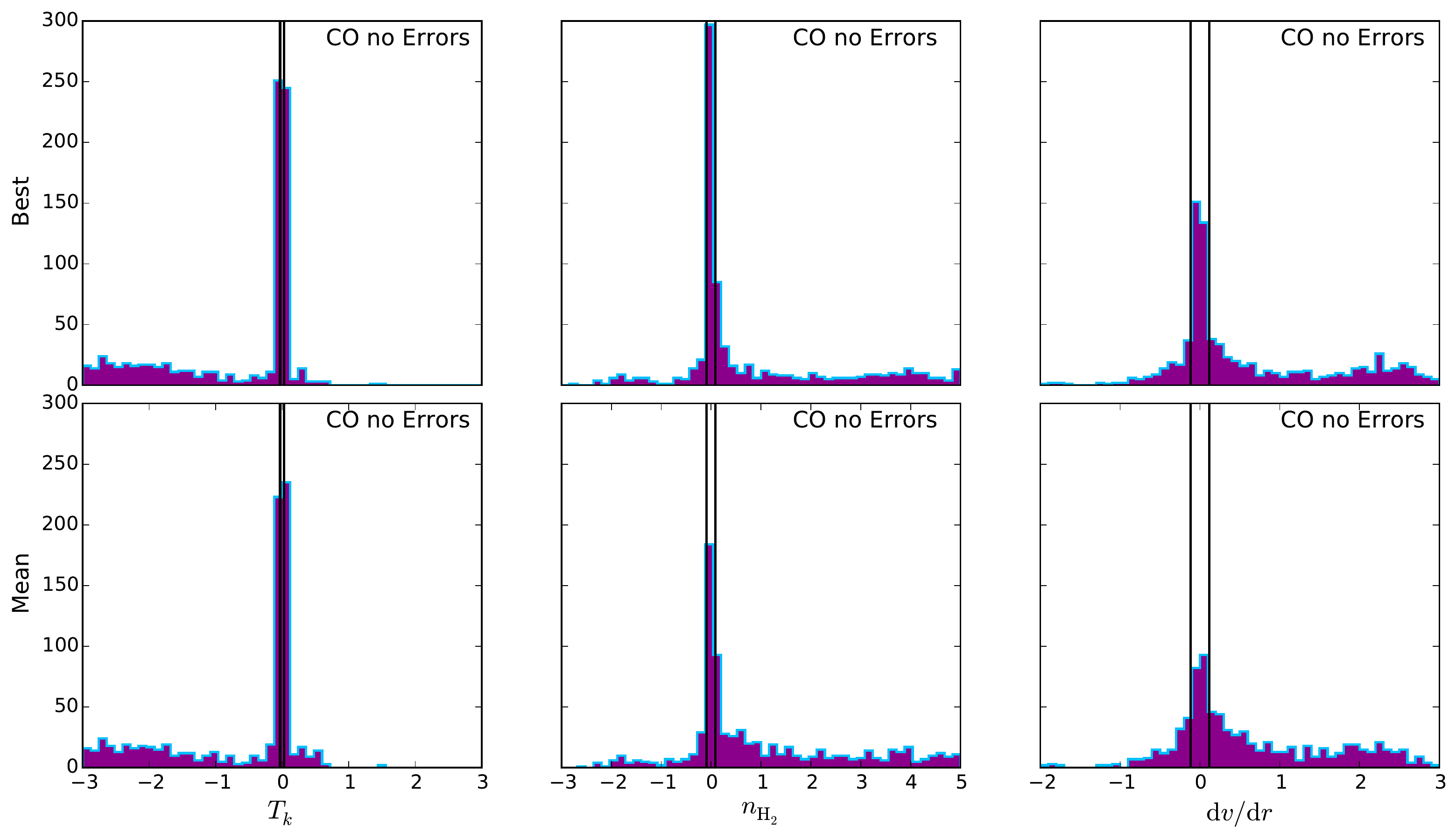}
\caption{The logarithm of the ratio of the recovered to true parameters with (upper panel) and without (lower panel) $10\%$ errors for CO, drawn from a pressure restricted parameter range and recovered with an MCMC. The best fit (top row) and mean (bottom row) recovered parameters are reported.}\label{fig:MCMCco}
\end{figure*}

\begin{figure*}
\centering
\includegraphics[width=\textwidth]{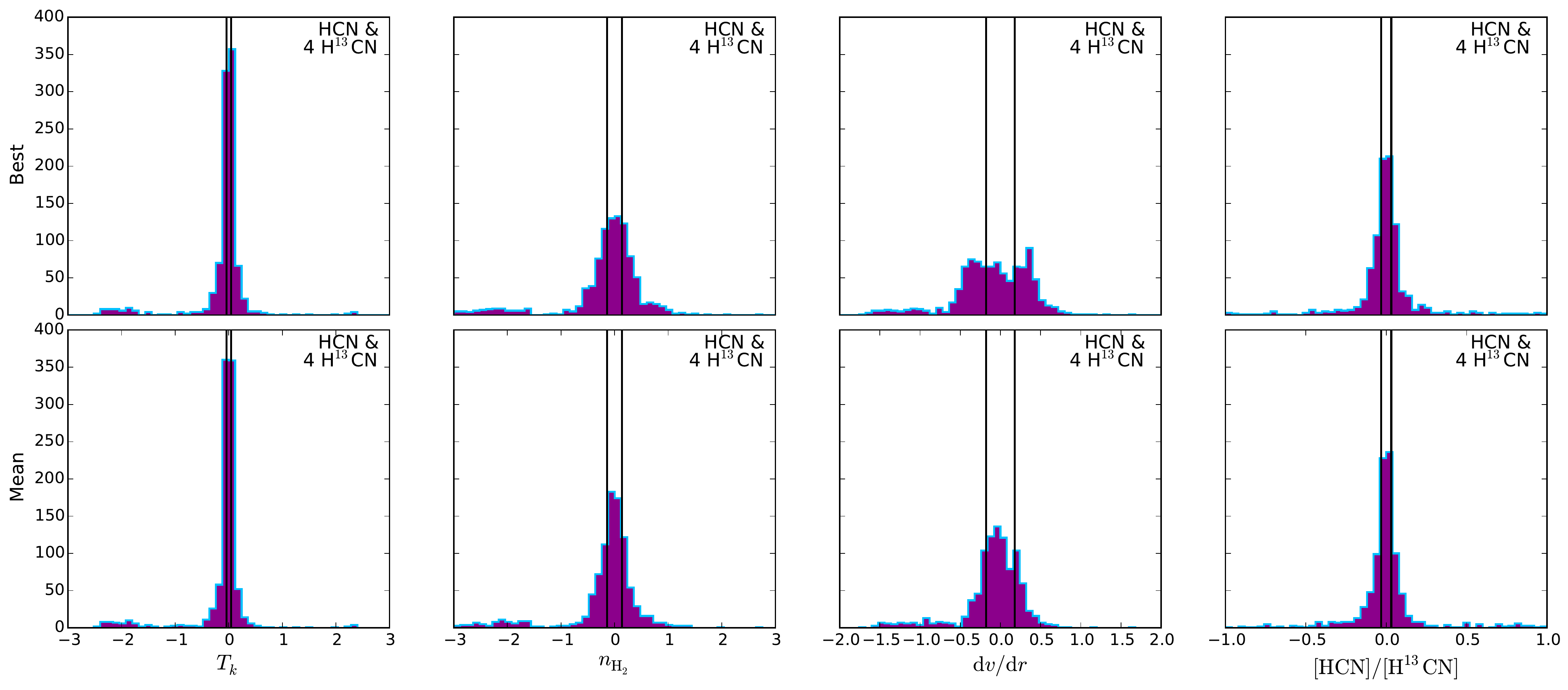}\\
\includegraphics[width=\textwidth]{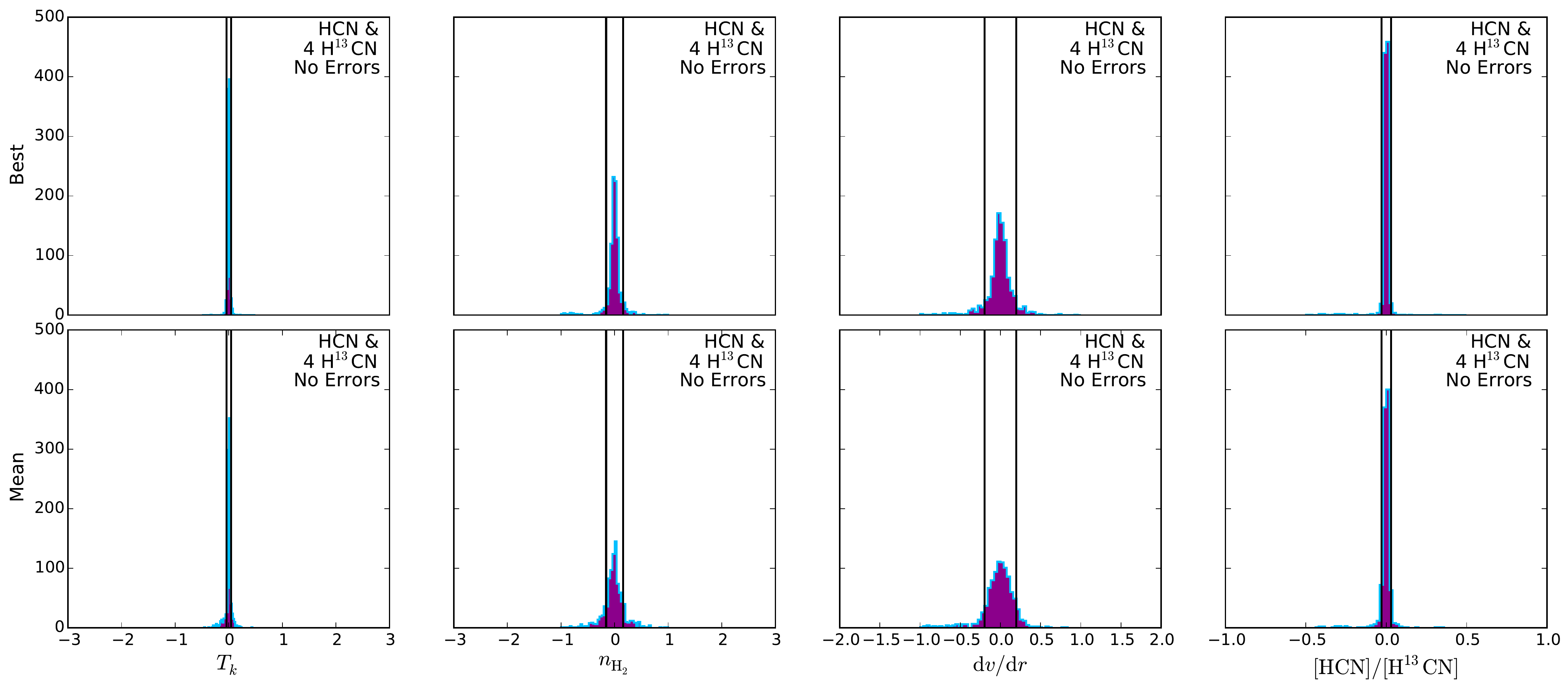}
\caption{The logarithm of the ratio of the recovered to true parameters with (upper panel) and without (lower panel) $10\%$ errors for four HCN lines and four H$^{13}$CN lines, drawn from a pressure restricted parameter range and recovered with an MCMC. The best fit (top) and mean (bottom) recovered parameters are reported. {The} $\pm1\sigma$ standard deviations of the MCMC traces are indicated by the vertical, solid black lines.}\label{fig:c12c13_4lines}
\end{figure*}

\begin{figure*}
\centering
\includegraphics[width=\textwidth]{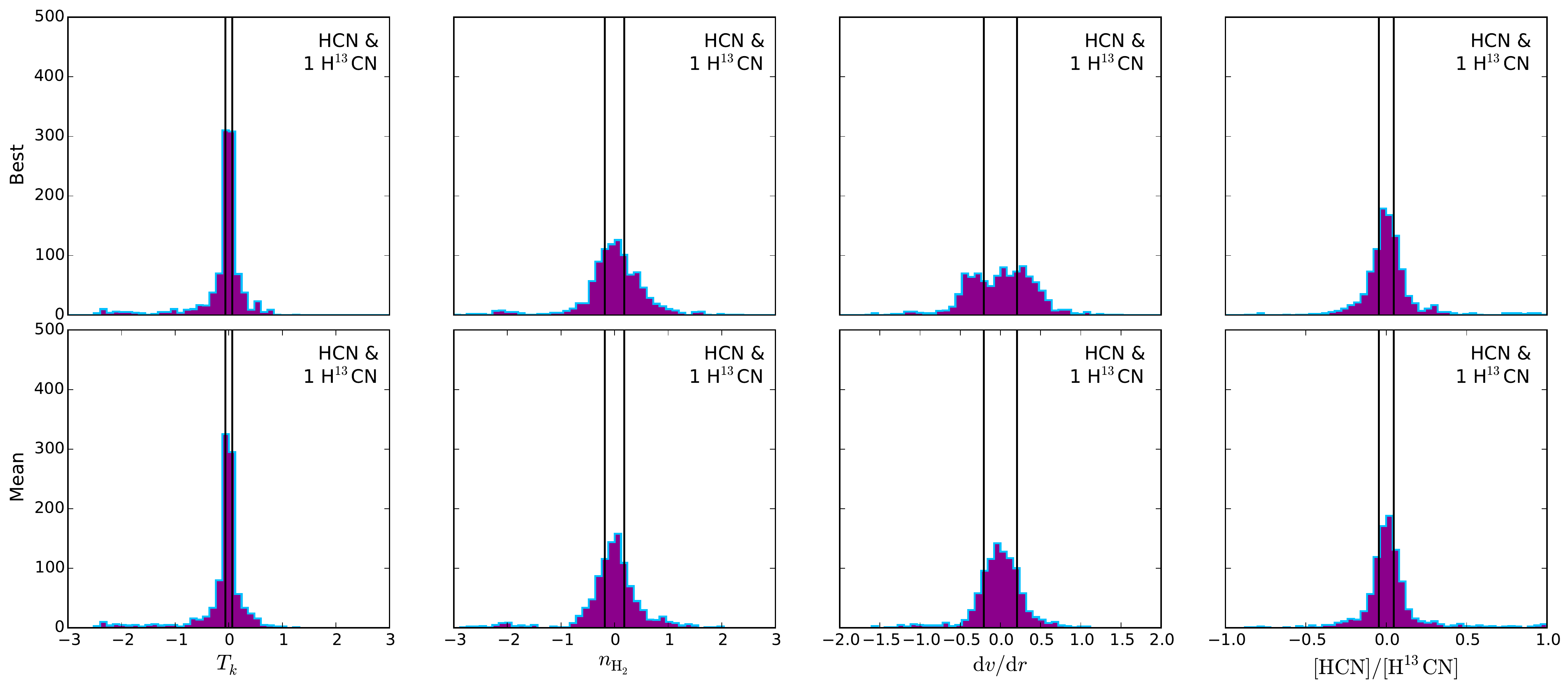}\\
\includegraphics[width=\textwidth]{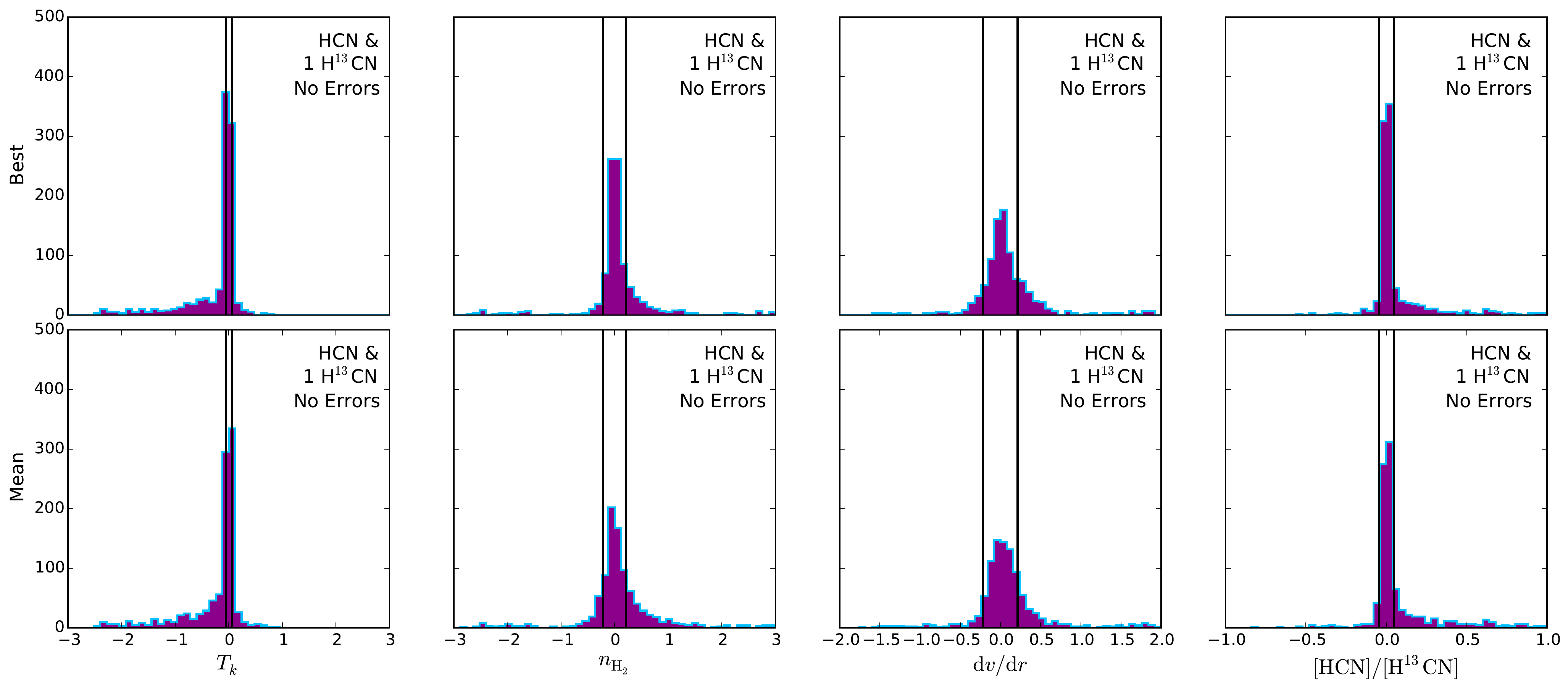}
\caption{The logarithm of the ratio of the recovered to true parameters with (upper panel) and without (lower panel) $10\%$ errors for four HCN lines and one H$^{13}$CN line, drawn from a pressure restricted parameter range and recovered with an MCMC. The best fit (top) and mean (bottom) recovered parameters are reported. {The} $\pm1\sigma$ standard deviations of the MCMC traces are indicated by the vertical, solid black lines.}\label{fig:c12c13_1line}
\end{figure*}

\begin{figure*}
\centering
\includegraphics[width=\textwidth]{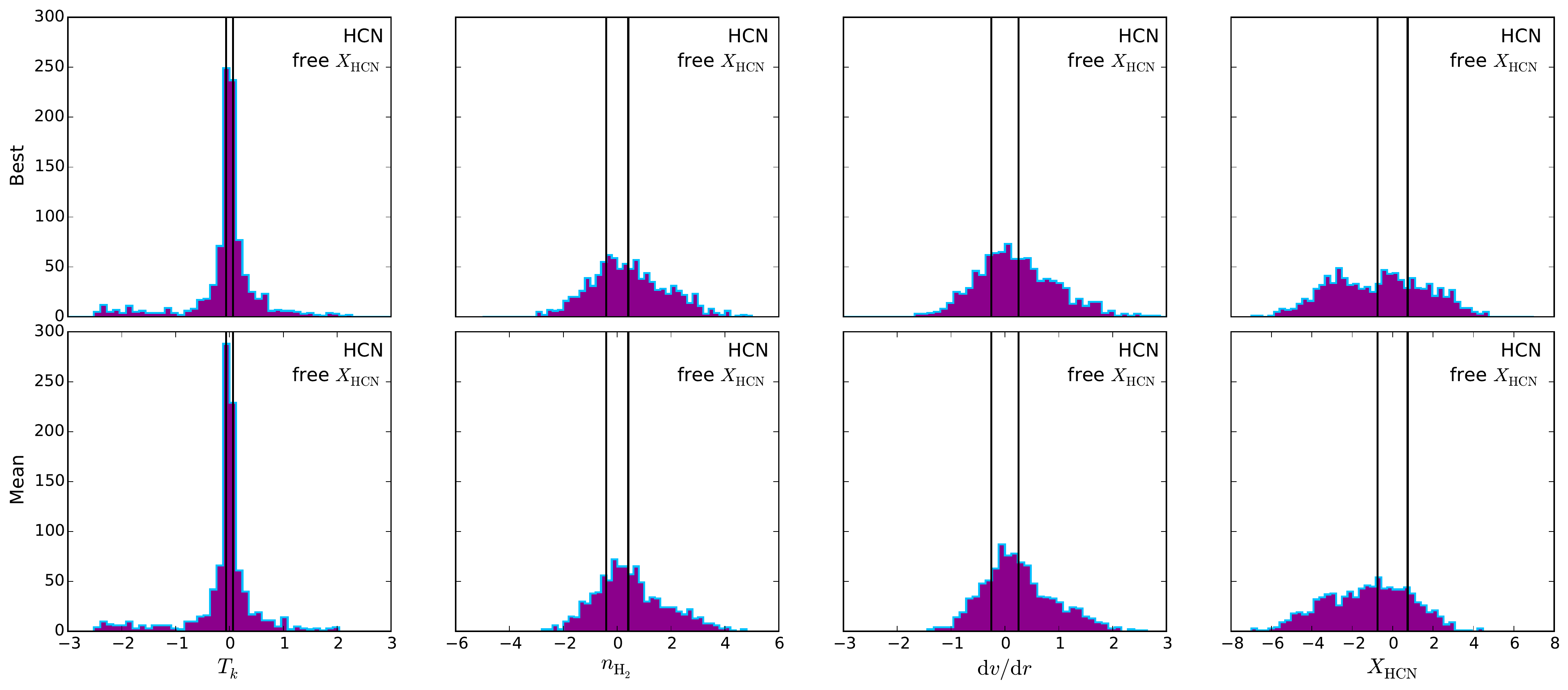}\\
\includegraphics[width=\textwidth]{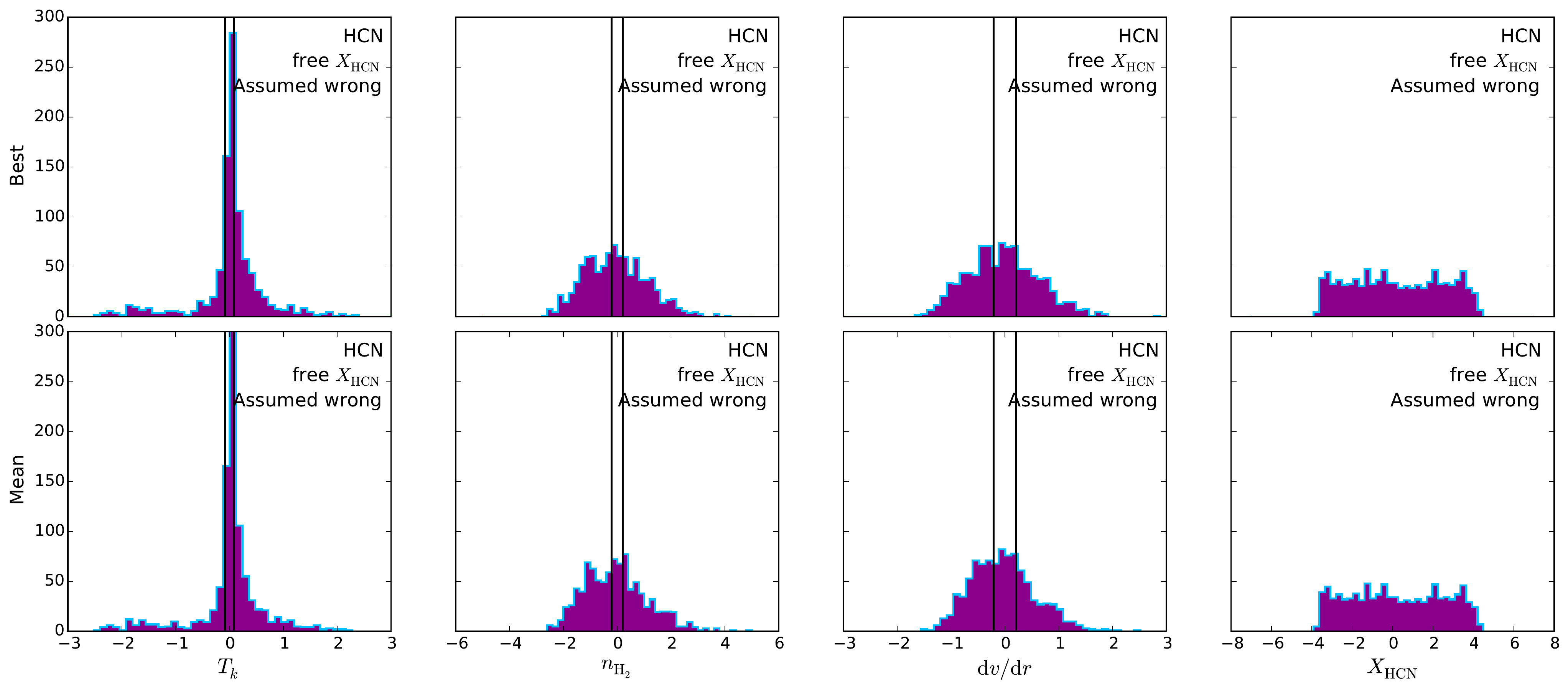}
\caption{The logarithm of the ratio of the recovered to true parameters with $10\%$ errors for four HCN lines, drawn from a pressure restricted parameter range and recovered with an MCMC. In all cases $X_{\rm HCN}$ is a random variable when generating the lines; in the top panel $X_{\rm HCN}$ is a free parameter in the recovery, while in the lower we attempt to recover parameters assuming $X_{\rm HCN}=2\times10^{-8}$. The best fit (top row) and mean (bottom row) recovered parameters are reported. {The} $\pm1\sigma$ standard deviations of the MCMC traces are indicated by the vertical, solid black lines, except for $X_{\rm HCN}$ which was a fixed parameter in the MCMC. $T_{\rm k}$ and $n_{{\rm H}_2}$ are recovered equally well in both cases.}\label{fig:fixXhcn}
\end{figure*}

\begin{figure*}
\centering
\includegraphics[width=\textwidth]{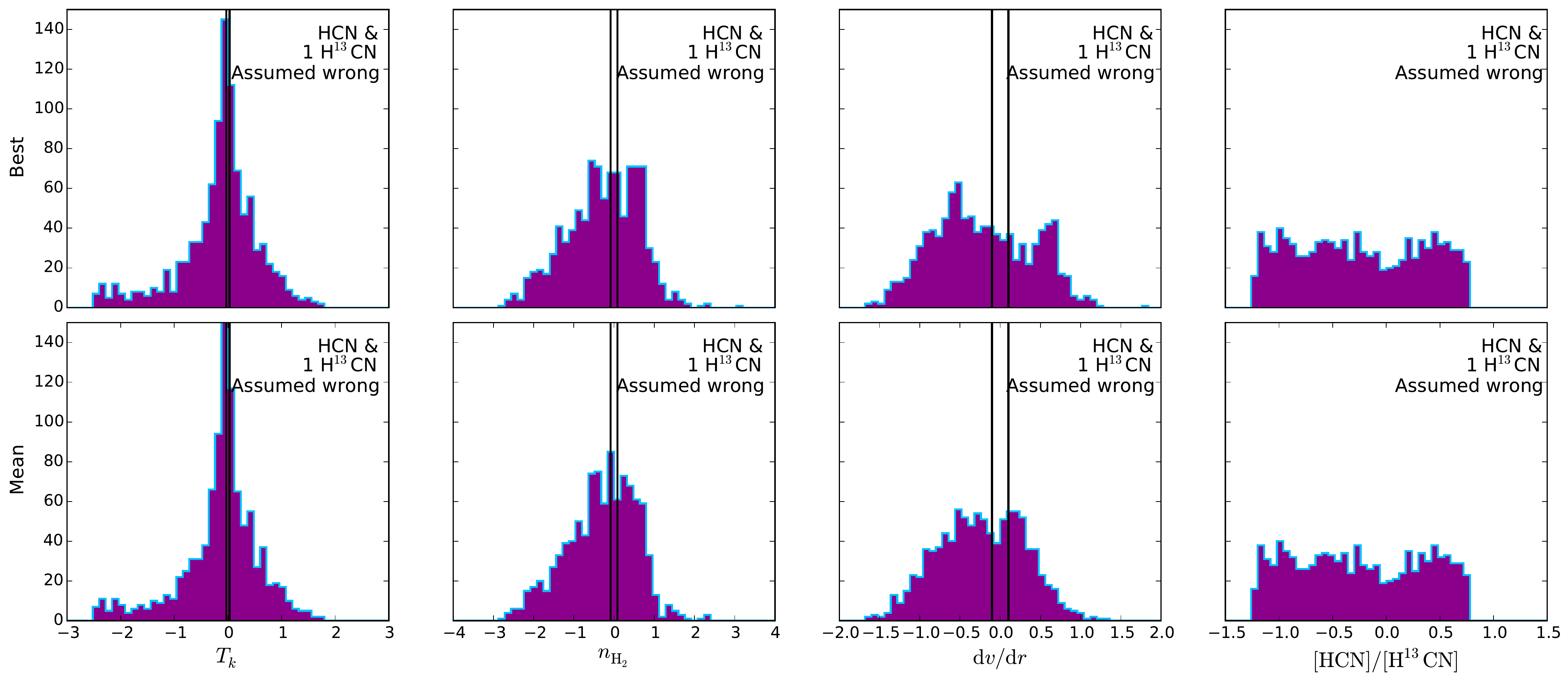}
\caption{The logarithm of the ratio of the recovered to true parameters with $10\%$ errors for four HCN lines and one H$^{13}$CN line, drawn from a pressure restricted parameter range and recovered with an MCMC, assuming [HCN]/[H$^{13}$CN]$ = 60$. The best fit (top) and mean (bottom) recovered parameters are reported. {The} $\pm1\sigma$ standard deviations of the MCMC traces are indicated by the vertical, solid black lines, except for [HCN]/[H$^{13}$CN] which was a fixed parameter in the MCMC.}\label{fig:fix13c}
\end{figure*}

\section{Free $X_{\rm mol}$}

We include a final model of less general interest, but which is nevertheless important for appreciating the limitations of LVG codes. We explore the case where we have four HCN lines and four H$^{13}$CN lines, and where both $X_{\rm mol}$ [HCN]/[H$^{13}$CN] are free parameters.

One of the key parameters necessary to predict the dominant heating source in a gas (PDRs, XDRs, MDRs etc.) is the molecular abundance. While a single molecular abundance is not particularly informative, with reliable estimate of the abundances of several species significant progress can be made \citep{Meijerink2005,Meijerink2007,Meijerink2011,Bayet2008,Viti2014}. An interesting question therefore is whether LVG models can in fact be used to reliably recover molecular abundances. While we showed in the main paper that with HCN lines alone this is not possible, we investigate whether the addition of H$^{13}$CN improves the situation.

Due to the degeneracy between $X_{\rm mol}$ and d$v/$d$r$ it is not possible to recover the molecular abundance with a single species, unless one is willing to assume a fixed value of $K_{\rm vir}$. If $K_{\rm vir}$ is assumed or known, then for a given $n_{{\rm H}_2}$ d$v/$d$r$ is determined. However, even this is not necessarily enough, due to the degeneracy between $T_{\rm k}$ and $n_{{\rm H}_2}$. Furthermore, the assumption of a fixed $K_{\rm vir}$ is dubious, especially in luminous infrared galaxies where $K_{\rm vir}$ is very likely elevated \citep{Papadopoulos2014}. This is before we take into account that even high resolution interferometric extragalactic observations (in all but the closest galaxies) will confuse multiple gas phases within the beam. 

\begin{figure*}
\centering
\includegraphics[width=\textwidth]{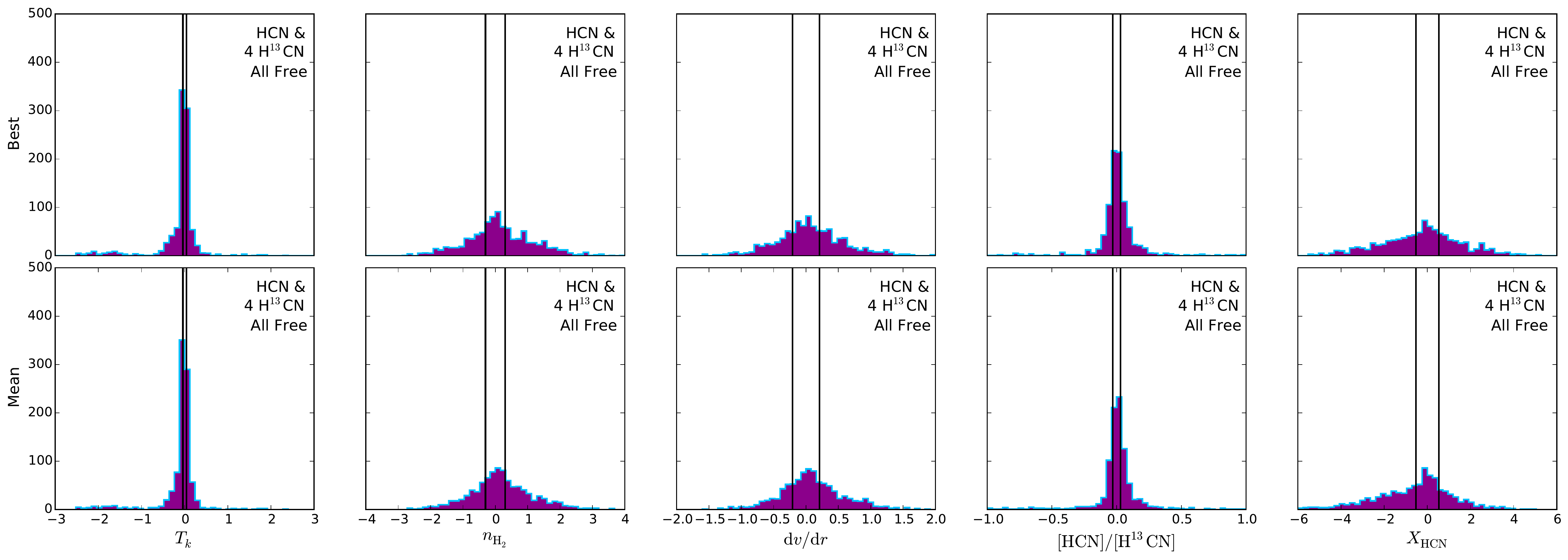}
\caption{The logarithm of the ratio of the recovered to true parameters with $10\%$ errors for four HCN lines and four H$^{13}$CN lines, drawn from a pressure restricted parameter range with free $X_{\rm HCN}$ and recovered with an MCMC. The best fit (top) and mean (bottom) recovered parameters are reported. {The} $\pm1\sigma$ standard deviations of the MCMC traces are indicated by the vertical, solid black lines. The model is not able to reliably recover the HCN abundance.}\label{fig:freehcn}
\end{figure*}

One possible solution appears to lie in isotopologue lines. Here, we present a model with four HCN lines and four H$^{13}$CN lines, with the free parameters $T_{\rm k}$, $n_{{\rm H}_2}$, d$v/$d$r$, $X_{\rm HCN}$ and [HCN]/[H$^{13}$CN]. The results are shown in Figure \ref{fig:freehcn}. Unfortunately, while $T_{\rm k}$ and [HCN]/[H$^{13}$CN] are well determined, $n_{{\rm H}_2}$, d$v/$d$r$ and $X_{\rm HCN}$ are not. The conclusion here is that while LVG models cannot reliably estimate molecular abundances, they \emph{can} very reliably estimate abundance ratios.

\section{CO Line Ratios}

\begin{figure*}
\centering
\includegraphics[width=\textwidth]{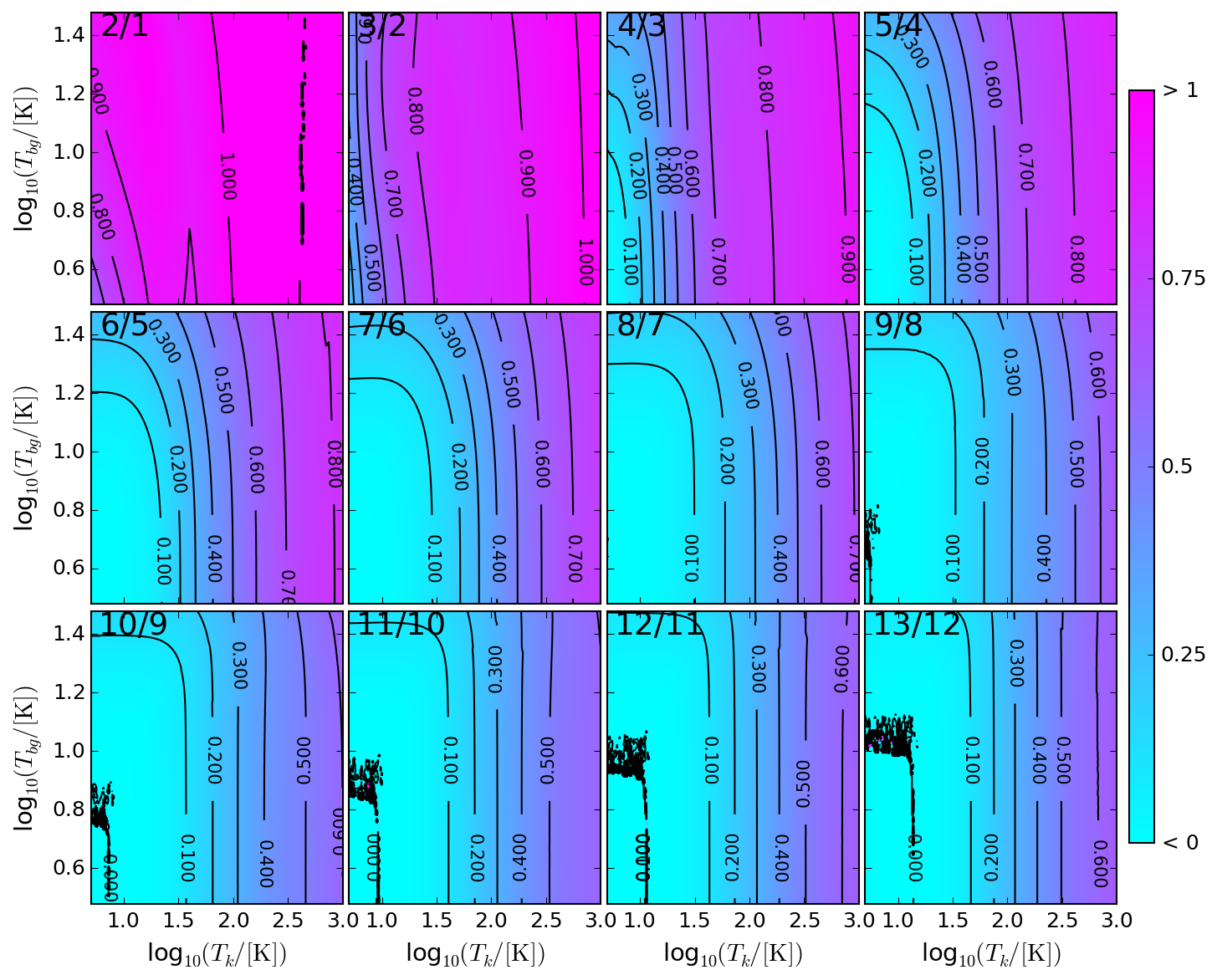}
\caption{The effect of increasing the background radiation temperature from 3\,K to 300\,K on a selection of CO line ratios for a range of kinetic temperatures from 10\,K to 1000\,K. The line ratio is given in the top right corner of each plot. We plot contours at ratio levels of $-10$, $-5$, $-2$, $-1$, $-0.5$, 0, 0.1, 0.2, 0.3, 0.4, 0.5, 0.6, 0.7, 0.8, 0.9, 1, 2, 5 and 10.}\label{fig:COtbgratios1}
\end{figure*}

\begin{figure*}[p]
\centering
\includegraphics[width=\textwidth]{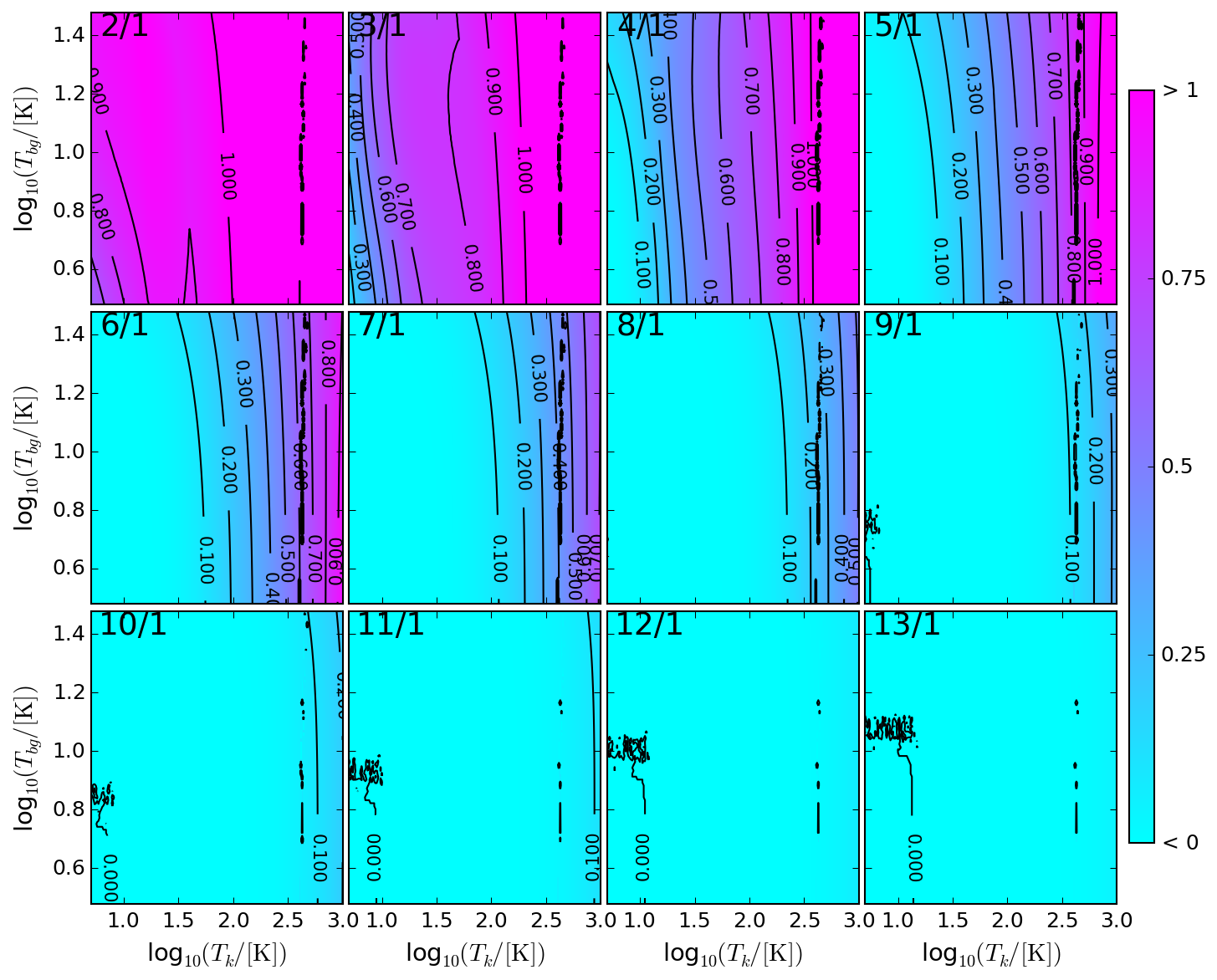}
\caption{{The effect of increasing the background radiation temperature from 3\,K to 300\,K on a selection of CO line ratios for a range of kinetic temperatures from 10\,K to 1000\,K. The line ratio is given in the top right corner of each plot. We plot contours at ratio levels of $-10$, $-5$, $-2$, $-1$, $-0.5$, 0, 0.1, 0.2, 0.3, 0.4, 0.5, 0.6, 0.7, 0.8, 0.9, 1, 2, 5 and 10.}}\label{fig:COtbgratios2}
\end{figure*}

As a continuation of the discussion of the effect of $T_{\rm bg}$ on the observed line fluxes and line brightness ratios in Section \ref{subsec:tbg_results} we present the CO line ratios in {Figures \ref{fig:COtbgratios1} and \ref{fig:COtbgratios2}}. It is not practical to present all 78 possible line ratios, so we include only the 12 ratios with the $J=1-0$ line and the 12 $J+1/J$ line ratios.


\end{document}